\documentclass[12pt]{article}
\usepackage{a4wide}
\usepackage{graphicx}
\usepackage[centertags]{amsmath}
\usepackage{amssymb}
\usepackage{cite}
\usepackage{enumerate}
\usepackage{scrtime}
\usepackage{color}
\usepackage[sans]{dsfont} 
\usepackage{colortbl}
\usepackage{subfigure}
\usepackage{paralist}
\usepackage{axodraw}
\usepackage{lscape}
\usepackage{longtable}
\usepackage{ulem}
\usepackage{cancel}

\graphicspath{{./}{./graphs/}}

\newcommand{\bs}[1]{\boldsymbol{#1}}
\newcommand{\bsb}[1]{\boldsymbol{\overline{#1}}}

\newcommand{\SO}[1]{\ensuremath{\mathrm{SO}(#1)}}
\newcommand{\SU}[1]{\ensuremath{\mathrm{SU}(#1)}}
\newcommand{\U}[1]{\ensuremath{\mathrm{U}(#1)}}
\newcommand{\E}[1]{\ensuremath{\mathrm{E}_{#1}}}

\newcommand{\gut}{{\textsc{gut}}}
\newcommand{\alphagut}{\alpha_{\gut}}

\newcommand{\alphaprime}{\alpha'}
\newcommand{\alphastring}{\alpha_{\textsc{string}}}

\newcommand{\alphaone}{\alpha_{\text{1}}}
\newcommand{\alphatwo}{\alpha_{\text{2}}}
\newcommand{\alphathree}{\alpha_{\text{3}}}
\newcommand{\ex}{{\textsc{ex}}}
\newcommand{\gev}{{\rm ~GeV~}}
\newcommand{\mssm}{{\textsc{mssm}}}
\newcommand{\mgut}{{M_{\gut}}}
\newcommand{\mex}{{M_{\textsc{ex}}}}
\newcommand{\mc}{{M_{\textsc{c}}}}
\newcommand{\mpl}{M_{\textsc{pl}}}
\newcommand{\mstring}{M_{\textsc{s}}}

\newcommand{\ptm}{\phantom{\text{-}}}

\def\z6ii{$\mathbb{Z}_6$-II}

\def\beqn{\begin{eqnarray}}
\def\eeqn{\end{eqnarray}}
\def\bsqn{\begin{subequations}}
\def\esqn{\end{subequations}}

\def\bctr{\begin{center}}
\def\ectr{\end{center}}

\def\mbsix{\mathbf{6}}
\def\mbsixb{\mathbf{\overline{6}}}
\def\mbthree{\mathbf{3}}
\def\mbthreeb{\mathbf{\overline{3}}}
\def\mbtwo{\mathbf{2}}

\def\st{{\textsc{string}}}
\begin{document}

\thispagestyle{empty}

\begin{flushright}
OHSTPY-HEP-T-08-002 \\
\end{flushright}
\vskip 2cm

\begin{center}
{\huge Reconciling Grand Unification with Strings\\[1ex]
by Anisotropic Compactifications\\[2ex]}
\vspace*{5mm} \vspace*{1cm}
\end{center}
\vspace*{5mm} \noindent
\vskip 0.5cm
\centerline{\bf Ben Dundee, Stuart Raby, Ak\i{}n Wingerter}
\vskip 1cm
\centerline{
\em Department of Physics, The Ohio State University,}
\centerline{\em 191 W.~Woodruff Ave, Columbus, OH 43210, USA}
\vskip2cm

\centerline{\bf Abstract}
\vskip .3cm

We analyze gauge coupling unification in the context of heterotic
strings on anisotropic orbifolds.  This construction is very much
analogous to effective 5 dimensional orbifold GUT field theories.
Our analysis assumes three fundamental scales,  the string scale,
$\mstring$, a compactification scale, $\mc$, and a mass scale for
some of the vector-like exotics,  $\mex$; the other exotics are
assumed to get mass at $\mstring$.  In the particular models
analyzed, we show that gauge coupling unification is not possible
with $\mex  = \mc$ and in fact we require $\mex \ll \mc \sim 3
\times 10^{16}$ GeV. We find that about 10\% of the parameter space
has a proton lifetime (from dimension 6 gauge exchange) $10^{33} \;
{\rm yr~} \lesssim\tau(p\rightarrow \pi^0e^+) \lesssim 10^{36}
{\rm ~yr}$. The other 80\% of the parameter space gives proton
lifetimes below Super-K bounds.  The next generation of proton decay
experiments should be sensitive to the remaining parameter space.

\vskip .3cm

\newpage

\section{Introduction}

Supersymmetric grand unification \cite{Dimopoulos:1981yj,
Dimopoulos:1981zb, Ibanez:1981yh, Sakai:1981gr, Einhorn:1981sx,
Marciano:1981un} is one of the most attractive scenarios for beyond
the Standard Model physics. One can simultaneously explain the
apparent unification of the electroweak and strong coupling
constants around $3 \times 10^{16} {\rm ~GeV}$, charge quantization,
the conservation of B-L, and why quarks and
leptons come in families.   Nevertheless the simplest four
dimensional SUSY GUTs have some notable problems.   Spontaneously
breaking the GUT symmetry requires scalars in adjoint
representations and complicated symmetry breaking potentials.  In
addition,  Higgs doublet-triplet splitting demands special
treatment.  Neither of these problems is insurmountable but it is
difficult to imagine that these special sectors can be derived from
a more fundamental theory.   In addition, Super-K bounds on the
proton lifetime place 4-dimensional SUSY GUTs ``under siege"
\cite{Dermisek:2000hr, Murayama:2001ur}.   Finally,  in order to
understand fermion masses and mixing angles it is likely that
additional family symmetries may be needed.

In the early work within the framework of the weakly coupled
heterotic string it was argued for string unification, as opposed to
grand unification with an independent lower energy GUT breaking
scale.\footnote{In fact, it is difficult to get massless adjoints in
the string spectrum, needed for GUT symmetry breaking.}  Gauge
couplings naturally unify at the string scale with a unification
scale\footnote{Assuming $SU(2)$ and $SU(3)$ are represented at
Ka\v{c}-Moody level $k_2=k_3=1$ and the $U(1)$ of hypercharge is
normalized with $k_1 = 5/3$.} of around $5 \times 10^{17} \gev$
\cite{Kaplunovsky:1987rp,Dixon:1989fj,Dixon:1990pc}. Unfortunately
the precision low energy data prefers a lower unification scale,
$\mgut \sim 3 \times 10^{16}$ GeV.  This tension between gravity and
gauge coupling unification has been termed the ``factor of 20"
problem with string unification \cite{Dienes:1996du}.  Nevertheless
string theory has some very nice features, i.e. the $\E{8} \times
\E{8}$ (or $\SO{32}$) symmetry of the weakly coupled heterotic
string is easily broken via an orbifold compactification of the
extra 6 spatial dimensions \cite{Dixon:1985jw,Dixon:1986jc}.   In
addition, Higgs doublet-triplet splitting is also easily
accomplished by the same means \cite{Breit:1985ud,Ibanez:1987sn}.
Significant progress was made early on in obtaining
standard-model-like theories using orbifolding and Wilson lines to
break the gauge symmetry
\cite{Ibanez:1986tp,Ibanez:1987xa,Ibanez:1987sn,Casas:1988hb,Casas:1987us}.

More recently, it was realized that some of the problems with SUSY
GUTS could be solved by understanding our low energy physics in
terms of an effective five or six dimensional field theory in which
one or two of the directions is compactified \cite{Dienes:1998vh,
Dienes:1998vg,Kawamura:2000ev,Hall:2001pg,Asaka:2001eh,Contino:2001si,Dermisek:2001hp,Hall:2002ea,
Kim:2002im,Lee:2006hwa}. Typically one takes a five (six)
dimensional gauge theory, and compactifies one (two) of the
directions on an orbifold. The geometry of the orbifold admits
solutions for higher dimensional fields which are localized on two
or more branes, and fields which are free to propagate in the bulk.
The former are called ``brane'' fields, the latter ``bulk'' fields.
By assigning the bulk fields boundary conditions along the fifth
(and sixth) direction(s), one can achieve GUT/SUSY breaking without
the large representations and complicated GUT breaking potentials
encountered in 4 dimensional constructions.  In addition, placing
the electroweak Higgs multiplet in the bulk, Higgs doublet-triplet
splitting can also be affected via a judicious choice of boundary
conditions. Generally, the placement of the matter and the
assignment of orbifold parities is done in a bottom up manner;  one
identifies certain phenomenological features (eg. suppressing
dangerous proton decay operators) and then chooses mass scales,
matter localization, and orbifold parities accordingly. For example,
one can keep $b-\tau$ Yukawa unification by placing the third family
on an \SU5 brane or suppress proton decay by placing the first two
families in the bulk \cite{Hall:2001pg}. Finally, four dimensional
SUSY GUTs require of order 3\% threshold correction at the GUT scale
in order to precisely fit the low energy data \cite{Yao:2006px}.
Given a GUT breaking sector, this correction must come from the
spectrum of massive states with mass of order $\mgut$. In orbifold
GUTs this correction comes from the Kaluza-Klein modes between the
compactification scale, $\mc$, and the cut-off scale, $M_*$, with
unification occurring at the cut-off. In fact, the ratio $M_*/\mc
\sim 100$ is determined by gauge coupling unification.  The problem
with orbifold GUT field theories, however, is the necessity for a
cut-off.

In Refs.~\cite{Kobayashi:2004ya,Forste:2004ie,Kobayashi:2004ud}, it
was shown that effective orbifold SUSY GUT field theories can be
obtained by orbifold compactifications of the heterotic string.
These theories provide an ultraviolet completion of orbifold GUT
field theories with a physical cut-off at the string scale.  These
are so-called anisotropic orbifold theories with one or two large
extra dimensions ($R = \mc^{-1} \gg l_{\textsc{S}} =
\mstring^{-1}$). At lowest order the gauge couplings unify at
$\mstring$.  Further, when working within the framework of the
weakly coupled heterotic string, there is a very specific
relationship between the strength of the GUT coupling and the
strength of gravity (see Equation (\ref{het_string_BC})).  Viewed in
this manner, the factor of 20 turns into a factor of 400 when
comparing to the (experimentally measured) value of Newton's
constant.  This makes it clear that there needs to be significant
threshold corrections (both logarithmic and power law) in order to
match the low energy data. In fact, important threshold corrections
are provided by Kaluza-Klein modes running in loops. Their spectrum
is calculable, and often gives non-trivial corrections to the
running of the couplings \cite{Dienes:1998vh, Dienes:1998vg}.

In this paper, we investigate ways to solve the ``factor of 20"
problem with heterotic string unification, within the context of the
orbifold GUT picture proposed in references
\cite{Kobayashi:2004ya,Forste:2004ie,Kobayashi:2004ud} \footnote{A
recent analysis of gauge coupling unification can also be found in
Reference \cite{Kim:2007jg}.}.  In order to make unification work,
we find that we generally need to introduce an intermediate scale,
$\mex$, which is typically two or three orders of magnitude below
the compactification scale.  When we impose the conditions that
$\mstring > \mc \gtrsim \mex$, we find a large number of solutions
for which unification works. Note the proton lifetime (from
dimension six operators) scales as the fourth power of $\mc$.  Most
solutions are excluded by proton decay, however a small number
predict proton lifetimes (from dimension six operators) that can be
measured in future experiments.

We begin with a brief review of the stringy embedding of orbifold
GUTs \cite{Kobayashi:2004ya,Forste:2004ie,Kobayashi:2004ud}, and a
presentation of the models in the ``mini-landscape" search
\cite{Buchmuller:2005jr, Buchmuller:2006ik,
Lebedev:2006kn, Lebedev:2006tr, Lebedev:2007hv} in Section
\ref{sec:ML}.  We focus on two ``benchmark" models from the
mini-landscape search in this analysis, called ``Model 1A'' and
``Model 2'' in Reference \cite{Lebedev:2007hv}. Specific details of
these models (the full spectrum in four dimensions, etc.) can be
found in Appendix \ref{app:model_two}. The main result of our
analysis is a detailed examination of the parameter space which
allows for unification, and how this parameter space relates to
proton decay constraints from dimension six (and possibly dimension
five) operators. This work is summarized in Section \ref{sec:GCU}.
Solutions consistent with gauge coupling unification are found in
Tables \ref{tab:proton_decay} - \ref{tab:interesting_models} on
pages \pageref{tab:proton_decay} - \pageref{tab:interesting_models}.
In Section \ref{sec:top_down} we check whether any of our solutions
are consistent with decoupling of exotics in supersymmetric vacua.

\section{\label{sec:ML}Orbifold GUTs from String Theory}

In exploring gauge coupling unification in orbifold constructions,
we focus on a class of models \cite{Buchmuller:2005jr, Buchmuller:2006ik,
Lebedev:2006kn, Lebedev:2006tr, Lebedev:2007hv} that are based on
$\SU6$ gauge-Higgs unification in five dimensions, and whose low-energy
spectrum is exactly that of the MSSM. Similar theories have also been
considered in the context of orbifold GUT field theory \cite{Hall:2001zb}.
We shall comment on the differences in Appendix \ref{sec:very_similar}.

\subsection{The Mini-Landscape in a Nutshell}
\label{sec:minilandscape-in-nutshell}

\begin{figure}[t!]
        \centering
        \includegraphics{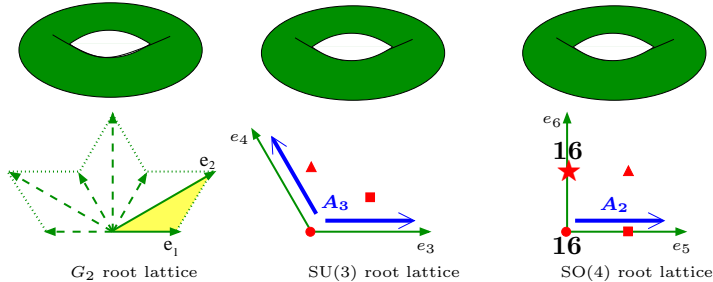}
        \caption{The geometry of the compact dimensions.}
        \label{fig:z6geometry}
\end{figure}

We compactify the 6 extra dimensions of the heterotic string on the product of three 2-tori
as shown in Figure \ref{fig:z6geometry}. Moding out the discrete $\mathbb{Z}_6$-II symmetry
given as a $60^\circ$, $120^\circ$, $180^\circ$ rotation (``twist'' $v$) in the first,
second, and third torus, respectively, defines the orbifold \cite{Dixon:1985jw,Dixon:1986jc}.
The geometry of the orbifold allows for no Wilson lines in the first torus, one order-3
Wilson line $A_3$ in the second torus ($e_3$ and $e_4$ are the same direction on the orbifold)
and two order-2 Wilson lines $A_2$, $A_2'$ along $e_5$, $e_6$, respectively \cite{Kobayashi:1991rp}.
We take $A_2' \equiv 0$ to localize 2 identical copies of $\bs{16}$'s at the fixed points
$\Red \bullet$ and $\Red \bigstar$ that will eventually sport a $D_4$ family symmetry
\cite{Kobayashi:2004ya,Kobayashi:2006wq,Ko:2007dz}.

Modular invariance allows for 61 different gauge embeddings
(``shift'' $V$) of the twist. Only 15 of these shifts break
$\E{8}\times\E{8}$ to a gauge group containing $\SO{10}$, and only 2
shifts allow for $\bs{16}$'s in the first/fifth twisted sector
($T_1$, $T_5$, respectively) that are not projected out by the
Wilson lines.

The models that come closest to the real world all stem from one
shift \cite{Buchmuller:2005jr, Buchmuller:2006ik}, termed $V^{\SO{10},1}$ in
Refs.~\cite{Lebedev:2006kn,Lebedev:2007hv}. Switching on all
possible Wilson lines consistent with this shift and modular
invariance, we obtain $\sim 22,000$ models with different particle
spectra. Successively, we impose our phenomenological priors to get
as close to the MSSM as we possibly can:
\begin{inparaenum}[(i)]
\item Standard Model gauge group,
\item non-anomalous hypercharge that lies in $\SU{5} \subset \SO{10}$,
\item 3 generations of quarks and leptons, 1 pair of Higgs doublets,
\item all exotic (i.e.~non-standard-model) particles are vector-like,
\item trilinear Yukawa coupling for a heavy top,
\item generalized B-L generator that is eventually broken down to $R$-parity,
\item all spurious abelian gauge group factors are broken,
\item string selection rules allow for all exotics to decouple consistent with the ``choice of vacuum''
(singlet VEVs must not break SM gauge symmetries and $R$-parity, and
must satisfy $F= D = 0$).
\end{inparaenum}

This leaves us with 15 models with promising phenomenology. We use
this sample to investigate whether the unification picture in
orbifolds is consistent with the measured values of the coupling
constants at low energies, or in other words, whether we can fit
$\alpha_1$, $\alpha_2$ and $\alpha_3$ at the electroweak scale with
a single coupling constant $\alphastring{}$ at $\mstring$.
Specifically, the set of exotics in both Models 1 and 2 of Reference
\cite{Lebedev:2007hv} are similar enough to warrant parallel
treatment, and are listed in Table \ref{tab:all_exotics} on page
\pageref{tab:all_exotics}. As can be seen, the exotic matter which
is charged under the MSSM in Model 1 overlaps with the exotic matter
in Model 2.  Note that we have labeled states with their hypercharge
and B-L quantum numbers as subscripts.

\subsection{The Orbifold GUT Picture}
\label{sec:orbifold-gut-picture}

The 15 models described in Section
\ref{sec:minilandscape-in-nutshell} are naturally embedded into a
grand unified theory in 5 or 6 dimensions \cite{Kobayashi:2004ud}.
Consider Model 2 of Section 5.2 of the mini-landscape search
\cite{Lebedev:2007hv}. For completeness, the full details of the
model have been reproduced in Appendix \ref{app:model_two}.

\begin{figure}[ht!]
\centering
\includegraphics{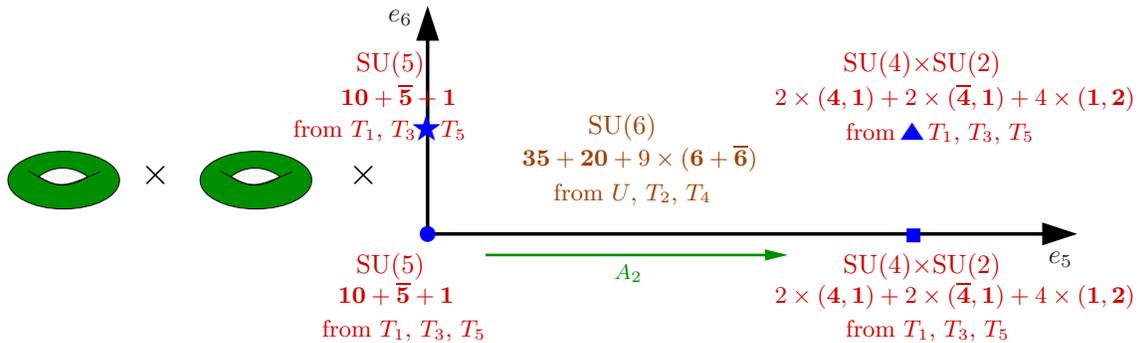}
\caption{Setup of the 5d orbifold GUT, where the 5th dimension ($e_5$) is large compared to the other compact dimensions.}
\label{fig:5d_picture}
\end{figure}

Instead of moding out the full \z6ii $\simeq
\mathbb{Z}_2\times\mathbb{Z}_3$ symmetry (generated by the twist
$v$) to get the 4-dimensional spectrum, we can mod out the
$\mathbb{Z}_3$ subgroup (generated by $2v$) alone, leaving the
$\SO{4}$ torus invariant. The particles from the $U$, $T_2$, $T_4$
sectors are free to move around in the $\SO{4}$ torus and can thus
be considered to be the ``bulk states'' of a 6-dimensional
$\mathbb{Z}_3$ orbifold with twist $2v$, shift $2V$ and Wilson line
$A_3$.

In this picture, moding out the residual $\mathbb{Z}_2$ symmetry
(generated by $3v$) corresponds to adding ``brane states'' to the
theory. The gauge group at the fixed points is obtained from the
bulk symmetry by moding out $V_2 = 3 V$ for $\Blue{\bullet}$ and
$\Blue{\bigstar}$ and $V_2 +A_2$ for $\Blue{\blacksquare}$ and
$\Blue{\blacktriangle}$. The matter representations follow from the
mass equation at the respective fixed points (given in terms of
$V_2$ and $A_2$), subject to projection conditions from $V_3 = 2V$
and $A_3$.

The gauge symmetry in 4 dimensions is the intersection of all gauge groups, and
the brane GUT states branch to SM representations of the $T_1$, $T_3$, $T_5$
sectors. This can be understood from an orbifold GUT viewpoint by assigning
parities to the brane modes given by
\begin{equation} \label{parities} \nonumber
P \sim e^{2\pi i p\cdot V_2} , \qquad P' \sim e^{2 \pi i p\cdot (V_2
+ W_2)},
\end{equation}
where $p$ (the highest weight associated with the state) is a
sixteen dimensional vector from the $\E8\times\E8$ lattice.  Then,
the setup of Figure \ref{fig:5d_picture} describes an orbifold
$S^1/\mathbb{Z}_2\times\mathbb{Z}'_2$ where 1 extra dimension is
compactified on a circle. The discrete symmetries are realized as a
reflection $\mathcal{P}:x^5\rightarrow -x^5$ and a translation
$\mathcal{T}:x^5\rightarrow x^5 + 2 \pi R$. Only the states that are
invariant under \beqn \label{GUT_parities} \nonumber
\mathcal{P}:\Phi(x^5)&\rightarrow& \Phi(-x^5) = P \Phi(x^5), \\
\mathcal{PT}:\Phi(x^5)&\rightarrow& \Phi(-x^5 + 2\pi R) = P'
\Phi(x^5) \eeqn will be present in the low energy
spectrum.\footnote{$P^\prime \equiv P \ T$ where $T$ corresponds the
discrete gauge transformation due to a Wilson line.}

Orbifold GUTs, when generated from an underlying string theory, are
significantly more constrained than orbifold GUT field theories.
Whereas the only \textit{real} constraint in an orbifold GUT field
theory is that the low energy effective field theory be anomaly
free, all anomalies in the string theory are canceled at the string
scale by the generalized Green-Schwarz mechanism
\cite{Green:1984sg,Witten:1984dg,Dine:1987xk,Sagnotti:1992qw,Berkooz:1996iz},
so this condition is automatically satisfied. In string orbifolds,
the parities are realized in terms of Wilson lines that must satisfy
stringent modular invariance constraints, so we cannot simply assign
parities at will.  Further, the placement of matter is not an
independent degree of freedom in string models.  Finally, we are
given a value for the coupling constant at the cut-off, see Equation
(\ref{het_string_BC}) on page \pageref{het_string_BC}. In a typical
orbifold GUT, this is a free parameter.  In addition, there may be
some assumptions about strong coupling, but the details of the
ultraviolet completion are not addressed.

\section{\label{sec:GCU}Gauge Coupling Unification in Orbifolds}

\subsection{Unification in Heterotic String Theory in 10 Dimensions}

As a unified framework for particle physics and gravity, string
theory predicts Newton's constant $G_N$ and relates it to the gauge
coupling constants. Unfortunately, the predicted value for $G_N$, in
the weakly coupled heterotic string, turns out to be too large and
needs to be reconciled with the extrapolated running gauge coupling
constants at the unification scale.

Throughout this paper we assume that we are in the weakly-coupled
regime of the heterotic string.  After compactifying the
10-dimensional low-energy effective action on a 6-dimensional
manifold, one obtains \cite{Kaplunovsky:1987rp} \beqn
\label{het_string_BC} G_N = \frac{1}{8}\,\alphastring\,\alphaprime.
\eeqn Here, $\alphastring$ denotes the common value of the gauge
coupling constants at the string scale, $\mstring =
1/\sqrt{\alphaprime}$.\footnote{The string scale $\mstring$ defined
here corresponds to the effective cut-off scale in our field theory
calculation.  This is discussed in more detail in footnote 8,
Section \ref{sec:anisotropic}.} Low-energy data suggests
$\alphastring^{-1} \simeq \alphagut^{-1} \simeq 24$ and $\mstring
\simeq \mgut \simeq 10^{16}$ GeV, so the predicted value for
Newton's constant is off by a factor of about 400. Putting it
another way, if we use the measured value of the gravitational
constant $G_N = 1/(\mpl^2)$ with $\mpl\simeq1.2\times10^{19} \gev$,
the string scale is predicted to be $\mstring \simeq 5\times10^{17}$
GeV \cite{Kaplunovsky:1987rp}, in disagreement with $\mgut$.  These
conclusions are based on the assumptions that
\begin{inparaenum}[(i)] \item we are in the weak coupling limit,
\item there are no new states between the electroweak and the GUT scale
that could contribute to the renormalization group equations (RGEs),
\item the compactification is isotropic, i.e.~all compactified dimensions are
comparable in size.\end{inparaenum}

In the following, we explore {\it anisotropic orbifold
compactifications} to fit low-energy data with a single coupling
constant at $\mstring$.\footnote{For earlier work along this line
see,
\cite{Ibanez:1991zv,Ibanez:1992hc,Kobayashi:2004ya,Hebecker:2004ce}.}
Other proposals that have been considered in the literature include
exotic matter representations at intermediate scales, large
threshold corrections, non-standard hypercharge normalizations from
higher-level Ka\v{c}-Moody algebras, strings without supersymmetry,
or the strong coupling regime of the heterotic string
\cite{Kaplunovsky:1987rp, Dixon:1989fj, Dixon:1990pc, string_unif,
Kim:2007jg}.  For a review of grand unification in the context of
string theory, see Reference \cite{Dienes:1996du}.

\subsection{The RGEs for Anisotropic Orbifold Compactifications}
\label{sec:anisotropic}

We study gauge coupling unification for the ``benchmark'' models
presented in the mini-landscape search \cite{Lebedev:2007hv}. As has
been emphasized in Section \ref{sec:minilandscape-in-nutshell},
these models are two out of 15 that already satisfy quite a few
non-trivial criteria on the road to the MSSM. We are working in the
the orbifold GUT limit as outlined in Section
\ref{sec:orbifold-gut-picture}. The gauge group geography and the
relevant part of its 5-dimensional spectrum for Model 2 are given in
Figure \ref{fig:5d_picture} on page \pageref{fig:5d_picture}. For
the full details of the 4-dimensional spectrum, see Tables
\ref{tab:spectrum_model1} and \ref{tab:spectrum_model2} on pages
\pageref{tab:spectrum_model1} and \pageref{tab:spectrum_model2}.
The anisotropic compactification singles out the fifth dimension
that is assumed to be large and thus introduces a new scale into the
theory, the compactification scale $\mc$. The other 5 compactified
dimensions are assumed to be of order the string scale, $\mstring$.

We want to compare our models with low energy data.   At the string
scale, $\mstring$, we have a unified gauge coupling, $\alphastring$.
Below the string scale we have three gauge couplings which
renormalize independently down to the weak scale. In general, there
are additional small (stringy) corrections to the relationship in
Equation (\ref{het_string_BC}) at the string scale, $\mstring$
\cite{Kaplunovsky:1987rp, Dixon:1989fj, Dixon:1990pc}. Because these
contributions are expected to be small, we will neglect them in this
analysis. In principle we should integrate the three gauge couplings
down to the SUSY breaking scale using the two-loop RGEs, including
one-loop threshold corrections at the string scale, the
compactification scale, the exotic scale, $\mex$, and the SUSY
scale, finally fitting $\alpha_i,i= 1,2,3$ at $M_Z$
\cite{Weinberg:1980wa, Hall:1980kf}. However, it is sufficient to
compare the orbifold GUT to the four dimensional SUSY GUT running
equations, which approximately (and implicitly) correct for SUSY
threshold corrections at the weak scale and two-loop renormalization
group running from the weak scale to the GUT scale. These are given
by the equations: \beqn \label{GQW_MSSM} \alpha_i^{-1}(\mu) =
\alphagut^{-1}+\frac{b_i}{2\pi} \log \frac{\mgut}{\mu}
+\frac{6}{2\pi} \delta_{i3}, \eeqn The indices $i=3,2,1$ refer to
$\SU{3}_c$, $\SU{2}_L$, $\U{1}_Y$, respectively. The $b_i$ are the
so-called $\beta$-function coefficients and are most conveniently
expressed in terms of the Dynkin index\footnote{For hypercharge, we
define the Dynkin index to be $\ell = (3/5) Y^2/4$.}
\cite{Slansky:1981yr}
\begin{equation}
b_i = -3 \ell(\text{vector multiplets}) + \ell(\text{chiral
multiplets}).
\end{equation}
For the MSSM we have $b_i = (-3, 1, 33/5)$.  Finally, the last term
in Equation (\ref{GQW_MSSM}) is a 3\% threshold correction to
$\alpha^{-1}_3$ at the GUT scale that we need to match the precision
electroweak data.

The minimal and most elegant way to fit the low-energy data is to
arrange for all exotics (i.e.~non-standard-model particles) to
obtain mass around $\mstring$. Up to the scale $\mc$, assumed to be
not much below $\mgut$, the evolution of the gauge coupling
constants is then governed by the same renormalization group
equations as in the usual GUT picture.  For energies above $\mc$,
the RGEs receive additional contributions from the Kaluza-Klein
tower of those Standard Model particles that live in the bulk, thus
giving rise to both logarithmic and power-law running
\cite{Dienes:1998vh, Dienes:1998vg}. Unfortunately, this simple
setup does not work. Varying the values of $\alphastring$ at
$\mstring$ and of the compactification scale $\mc$, we cannot fit
the gauge coupling constants at the electroweak scale.  We elaborate
on this point in Appendix \ref{sec:very_similar}, where we show the
difficulties involved with gauge-Higgs unification in five dimensions.

The remaining possibility is to assume that not all exotics obtain
mass at $\mstring$, but some are light enough to be relevant for the
evolution of the coupling constants. At the same time, of course,
the exotics must still be massive enough to decouple from the
low-energy theory. We will call this intermediate scale $\mex$ and
assume in the following $\mstring > \mc \gtrsim \mex$. Now we can
try to fit the low-energy data by varying $\mex$, $\mc$, $\mstring$
and the multiplicities and quantum numbers of the light exotics.
Note that the running of the coupling constants below $\mex$ will be
given by the same Equation (\ref{GQW_MSSM}) as in the MSSM, since all the
exotics are assumed to be heavier than $\mex$ and the first
excitation of the Kaluza-Klein tower is of order $\mc$. Near $\mu
\simeq \mex$, the renormalization group equations read: \beqn
\nonumber \label{pre_running_fiveD} \alpha_i^{-1}(\mu) &=&
\alphastring^{-1} +
\frac{b_i^{\mssm,++}+b_i^{\mssm,\text{brane}}}{2\pi} \log
\frac{\mstring}{\mu} \\ \nonumber && +
\frac{b^{\ex,++}_i+b_i^{\ex,\text{brane}}}{2\pi}\log\frac{\mstring}{\mex}
\\ \nonumber
&& -\frac{1}{4\pi} \left(b_i^{\mssm,++} + b_i^{\mssm,--} + b_i^{\ex,++} + b_i^{\ex,--}\right) \log \frac{\mstring}{\mc} \\
&& + \sum_{P=\pm,P'=\pm} \frac{b_i^{\mssm,PP'} + b_i^{\ex,PP'}}{2
\pi}\left(\frac{\mstring}{\mc} - 1\right) \eeqn These equations are
obtained by starting at the highest scale in the theory, $\mstring$,
and evolving the gauge couplings $\alpha_i$ down to $\mc$, taking
into account all the particles with mass less than $\mstring$. In
the next step, one takes the values obtained for $\alpha_i$ as
boundary conditions for the renormalization group equations at $\mc$
and calculates $\alpha_i$ at $\mex$. In order to compare to
experimental values of the coupling constants at $M_{\textsc{z}}$,
we apply the two loop RGEs \cite{Jones:1981we}. Note that this
involves integrating out SUSY particles at $M_{\textsc{susy}}$.
Technically, because the two loop RGEs are good \textit{near} the
GUT scale, our approach will be to compare the two equations
(Equations (\ref{GQW_MSSM}) and (\ref{pre_running_fiveD})) at the
scale $\mex$. Provided that $\mex$ is \textit{near} the GUT scale,
the error introduced in the analysis should be negligible. In
principle, the exotic scale $\mex$ can be small, perhaps a TeV.  In
all cases we find, however, the exotic scale is larger than $10^{9}
\gev$, and in most cases it is greater than $10^{12} \gev$.  The
error we make by matching Equations (\ref{GQW_MSSM}) and
(\ref{pre_running_fiveD}) at $\mex \sim 10^9$ GeV comes from the
difference in the two loop corrections to the RG running from $\mex$
to the GUT scale. This correction is expected to be less than a
percent.

Let us look at Equation (\ref{pre_running_fiveD}) in some more
detail. The first term is the tree level boundary condition from the
heterotic string.  The second and third terms contain loop
contributions from MSSM fields and exotic matter, respectively---the
zero KK modes and the brane states are kept separate for clarity.
The last two terms are due to the massive KK states in the bulk. The
logarithmic ($\sim \log \frac{\mstring}{\mc}$) and linear terms
($\sim \frac{\mstring}{\mc}$) are a consequence of the geometry,
i.e. in an equivalent string calculation the factor of
$\frac{\mstring}{\mc}$ arises from the dependence on the $T$
(volume) and $U$ (shape) moduli of the torus.\footnote{Note, our one
loop calculations are performed using an effective field theory
approach. In particular the sum over the infinite tower of KK modes
follows the regularization scheme of Dienes et al.
\cite{Dienes:1998vh,Dienes:1998vg}. Moreover, in the work of
Ghilencea and Nibbelink \cite{Ghilencea:2002gc} it is shown that if
the field theory cut-off $\Lambda^2$ is chosen to satisfy the
relation $\Lambda^2 = \frac{2 e}{3 \sqrt{3}} \frac{1}{\alpha^\prime}
\approx 1.05/\alpha^\prime$ then the heterotic string loop
calculation is approximately equal to the field theory results. Thus
we identify the string scale $\mstring = \Lambda \approx
\frac{1}{\sqrt{\alpha^\prime}}$. We should note that the analysis of
\cite{Ghilencea:2002gc} was done in the context of toroidal
compactification.  A more relevant comparison should be done in an
orbifold compactification with Wilson lines. The latter approach was
taken by the authors of Reference \cite{Kim:2007jg} in a
$T^2/\mathbb{Z}_3$ orbifold. Their results, however, are not
directly applicable to our situation.} Note, the last term is a
universal correction due to the \SU6 fields in the bulk. We
introduce the following definitions:
\begin{gather}
b_i^{\mssm} \equiv b_i^{\mssm,++} + b_i^{\mssm,\text{brane}}, \qquad b_i^{\ex} \equiv b^{\ex,++}_i+b_i^{\ex,\text{brane}}, \qquad b_i^{++} \equiv b_i^{\mssm,++} + b_i^{\ex,++} \nonumber \\
b_i^{--} \equiv b_i^{\mssm,--} + b_i^{\ex,--}, \qquad b^{\mathcal{G}} \equiv \sum_{P=\pm,P'=\pm} b_i^{\mssm,PP'} + b_i^{\ex,PP'} \nonumber
\end{gather}
This simplifies Equation (\ref{pre_running_fiveD}) a bit:
\beqn \nonumber \label{running_fiveD}
\alpha_i^{-1}(\mu) &=& \alpha_{\st}^{-1} + \frac{b_i^{\mssm}}{2\pi} \log \frac{\mstring}{\mu} + \frac{b_i^{\ex}}{2\pi}\log\frac{\mstring}{\mex} \\
&& -\frac{1}{4\pi} \left(b_i^{++} + b_i^{--}\right) \log
\frac{\mstring}{\mc} + \frac{b^{\mathcal{G}}}{2
\pi}\left(\frac{\mstring}{\mc} - 1\right) \eeqn

\subsection{Gauge Coupling Unification: \\ An Effective Field Theory Calculation}

Before we proceed, we will clear up some notational issues.  We will
\textit{always} talk about fields in the language of $N=1$ SUSY in
four dimensions. The $N$ = 1, 5-dimensional hyper multiplet contains two
4-dimensional chiral multiplets, and a 5-dimensional vector
multiplet contains a 4-dimensional vector multiplet and a
4-dimensional chiral multiplet. The 5-dimensional $N=1$ theory can
thus be described in terms of 4-dimensional $N=1$ fields (or in
terms of 4-dimensional $N$=2 hyper multiplets).

In order to check gauge coupling unification, we will equate the values of \begin{inparaenum}[(i)]
\item $1/\alphathree-1/\alphatwo$, \item $1/\alphatwo-1/\alphaone$, \item $\alphathree$  \end{inparaenum}
as obtained from Equation (\ref{GQW_MSSM}) and Equation (\ref{running_fiveD}),
respectively, at the scale $M_{\ex}$, where both equations are
valid. We find: \bsqn \label{diff_running} \beqn \label{first_diff}
\log\frac{\mstring}{\mgut} &=& \frac{n_3-n_2}{4} \log\frac{\mstring}{\mex} - \frac{3}{2}, \\
\label{second_diff}
\log\frac{\mstring}{\mgut} &=& \frac{10n_2-n_3-3n_1}{56} \log\frac{\mstring}{\mex} + \frac{3}{7} \log\frac{\mstring}{\mc}, \\ \nonumber
48\pi &=& \frac{\pi}{4}\left(\frac{\mpl}{\mstring}\right)^2 - 6 -3\log\frac{\mstring}{\mgut} + n_3\log\frac{\mstring}{\mex} \\
&&+ \log\frac{\mstring}{\mc}- 4\left(\frac{\mstring}{\mc}-1\right),
\eeqn \esqn where the $n_i$ are beta function contributions from the
\textit{brane localized} exotics, as defined below. The first two
equations describe the \textit{relative} running of the couplings
(i.e.~their slopes), and the last one gives us information about the
\textit{absolute} running (i.e.~their intercepts).  The coefficients
$n_i$ are defined in terms of the set of exotics with mass of order
$\mex$ as follows: \beqn \label{define_n_values} n_3\times
\left[(\mbthree,1)_{1/3,*}+(\overline{\mbthree},1)_{-1/3,*}\right] +
n_2 \times \left[(1,\mbtwo)_{0,*}+(1,\mbtwo)_{0,*}\right] + n_1
\times \left[(1,1)_{1,*}+(1,1)_{-1,*}\right], \eeqn where ``$*$''
for the B-L charge denotes anything.  The necessary $\beta$-function
coefficients $b_i$, using the numbers in Tables
\ref{tab:vector_like_brane} on page \pageref{tab:vector_like_brane}
are found to be
 \beqn \label{b_values}
 \vec{b}_{\ex} = (n_3,n_2,\frac{n_3+3n_1}{10}).
\eeqn

Let us now consider those MSSM states located in the bulk. In
general, we can find two pairs of $N=1$ chiral multiplets
$\mathbf{6} + \mathbf{6}^c$ which decompose as \beqn \nonumber 2
\times (\mbsix+\mbsix^c) &\supset&
\left[(1,\mbtwo)_{1,1}^{--}+(\mbthree,1)_{-2/3,1/3}^{-+}\right]+
\left[(1,\mbtwo)_{-1,-1}^{++}+(\overline{\mbthree},1)_{2/3,-1/3}^{--}\right]
\\
&&+\left[(1,\mbtwo)_{1,1}^{-+}+(\mbthree,1)_{-2/3,1/3}^{--}\right]+
\left[(1,\mbtwo)_{-1,-1}^{+-}+(\overline{\mbthree},1)_{2/3,-1/3}^{++}\right].
\eeqn This gives us the the third family $\bar b$ and $L$---the rest
of the third family comes from the $\mathbf{10} + \mathbf{10}^c$ of
\SU5 contained in the $\mathbf{20} + \mathbf{20}^c$ of \SU6, which
lives in the untwisted sector. An interesting point is the genesis
of the Higgs bosons. We have remarked earlier that the models we
look at come from a broader class of models satisfying ``gauge-Higgs
unification''.  Our bulk gauge symmetry is $\SU{6}$, so the $\SU{6}$
gauge bosons (and thus the adjoint representation) necessarily live
in the bulk.  Under $\SU{5} \times \U{1}$, the adjoint decomposes as
\begin{equation}
   \mathbf{35} \rightarrow \mathbf{24}_0 + \mathbf{5}_{+1} + \mathbf{5}^c_{-1} + 1_0.
\end{equation}
Thus the MSSM Higgs sector emerges from the breaking of the $\SU{6}$ adjoint by the orbifold.
Including the contributions from the third family and the Higgses, we find using Table
\ref{tab:all_b_values} on page \pageref{tab:all_b_values}
\begin{equation}
 \vec{b}^{++} = (-7,-3,13/5), \quad \vec{b}^{--} = (5,1,1/5), \qquad b^{\mathcal{G}} = -4.
\end{equation}

\subsection{Results}

\begin{figure}[ht!]
        \centering
        \includegraphics[scale = 0.5]{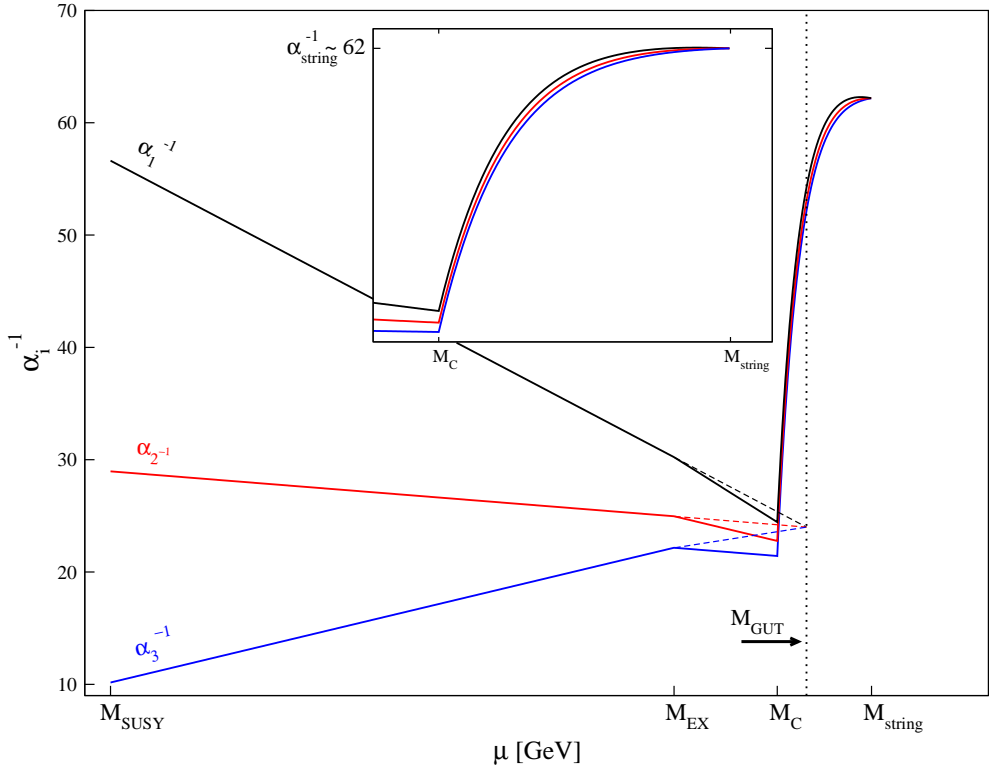}
        \caption{An example of the type of gauge coupling evolution we see in these models,
        versus the typical behavior in the MSSM.  The ``tail'' is due to the power law running
        of the couplings when towers of Kaluza-Klein modes are involved.  Unification in this
        model occurs at $\mstring \simeq 5.5\times 10^{17} \gev$, with a compactification scale
        of $\mc \simeq 8.2 \times 10^{15} \gev$, and an exotic mass scale of $\mex \simeq 8.2 \times 10^{13} \gev$.}
        \label{fig:beta_functions}
\end{figure}

We find it necessary to introduce an intermediate mass scale $\mex$,
perhaps \textit{near} the compactification scale, and identify a set
of exotics with mass $\mex$ consistent with gauge coupling
unification.  Solving the RG equations numerically, we find 252
versions of Model 2 (of which 82 are also versions of Model 1),
where by ``versions" we mean inequivalent sets of ``light" exotics
satisfying gauge coupling unification.  Of these 252 (82), only 48
(9) are consistent with the Super-K bounds on the proton lifetime
\cite{Yao:2006px} (see Section 3.5). These are found in Tables
\ref{tab:proton_decay} and ~\ref{tab:proton_decay_two} on pages
\pageref{tab:proton_decay} and \pageref{tab:proton_decay_two}, where
we also calculate the lifetime of the proton due to dimension six
operators, see Appendix \ref{sec:proton-decay} and Figure
\ref{fig:histogram}. The solutions which are applicable to Model 1
are listed in \textbf{bold} in both tables. Note that the GUT
coupling constant, $\alphastring$, (evaluated at $\mstring$) varies
depending on $\mstring$ and $\mex$. For example, in the last row of
Table \ref{tab:proton_decay} on page \pageref{tab:proton_decay}, we
find \beqn \alphastring^{-1} = \frac{1}{8}
\left(\frac{\mpl}{\mstring}\right)^2 \simeq
\frac{1}{8}\left(\frac{1.22\times 10^{19}\gev}{5.47 \times
10^{17}\gev}\right)^2 \simeq 62. \eeqn Near the exotic scale where
we match onto the low energy physics, we expect the (inverse)
coupling constants to be of order 30-40.  Likewise,
$\alphastring^{-1}$ is typically \textit{larger} than this, of order
50-60 or so (but sometimes as big as $\mathcal{O}(1000)$). Thus, we
\textit{must} have a large and negative contribution from the
power-law running, which translates into the requirement that
$b^{\mathcal{G}} < 0$.  This is evident in Equation
(\ref{running_fiveD}), for example.  If $b^{\mathcal{G}} > 0$, we
would need a large negative contribution from the other terms, which
is hard to reconcile with the logarithmic suppression. For
completeness, we plot the $\beta$-functions of the last solution in
Table \ref{tab:proton_decay} in Figure \ref{fig:beta_functions}. The
evolution of the gauge couplings is typical in this class of models,
i.e. the power law running between the compactification scale is
rather pronounced.

\begin{figure}[t!]
        \centering
        \includegraphics[scale = 0.5]{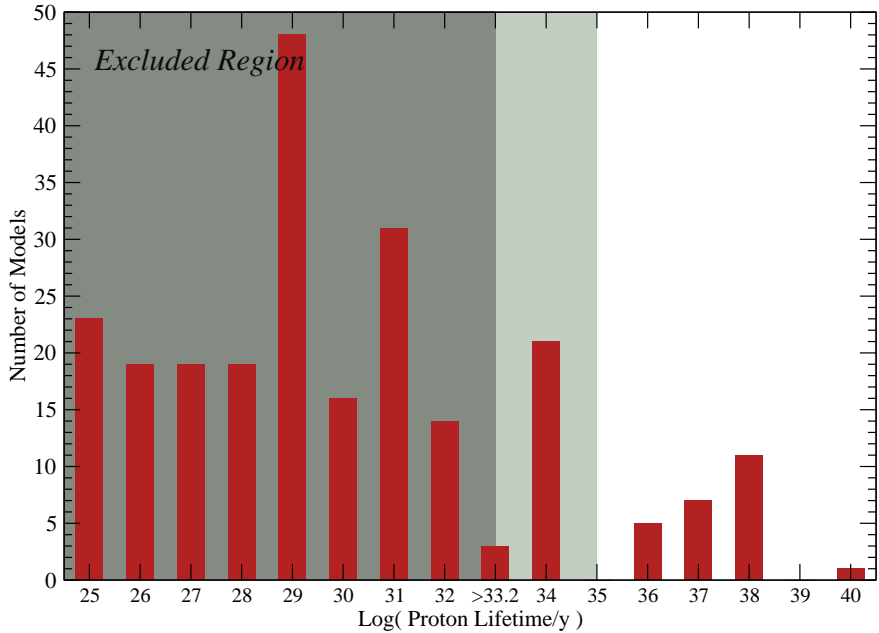}
        \caption{Histogram of solutions with $\mstring > \mc \gtrsim \mex$, showing the
        models which are excluded by Super-K bounds (darker green) and those which are
        potentially accessible in a next generation proton decay experiment (lighter green).
        Of 252 total solutions, 48 are not experimentally ruled out by the current
        experimental bound, and most of the remaining parameter space can be eliminated
        in the next generation of proposed proton decay searches.}
        \label{fig:histogram}
\end{figure}

For the 11 models in Table \ref{tab:proton_decay}, we keep only the
minimum amount of matter in the bulk, i.e.  in order to get the MSSM
spectrum, it is sufficient to keep $2 \times (\mathbf{6} +
\mathbf{6}^c)$ massless below the string scale.  Given the
constraint that we want $b^{\mathcal{G}} < 0$, however, we are in
principle able to leave $4 \times (\mathbf{6} + \mathbf{6}^c)$
massless below the string scale. This gives $b^{\mathcal{G}} = -2$,
and leads to 37 new solutions. These are listed in
Table \ref{tab:proton_decay_two} on page \pageref{tab:proton_decay_two}.
Of the 48 solutions (included in
both Models 1 and 2), 22 have proton lifetimes which can potentially
be tested by the next generation of proton decay experiments, see
Appendix \ref{sec:proton-decay} and Figure \ref{fig:histogram} for more
details.

We stress that this analysis is quite
general. Of the fifteen models which fit the criteria in the mini-landscape search,
all come from a five dimensional \SU6 orbifold, and all of them have the same types
of exotics.  This means that the analysis preformed here generalizes in a straightforward
manner to the other min-landscape models, whose spectra are listed in Reference \cite{tables}.

\begin{figure}[t!]
        \centering
        \includegraphics[scale = 0.5]{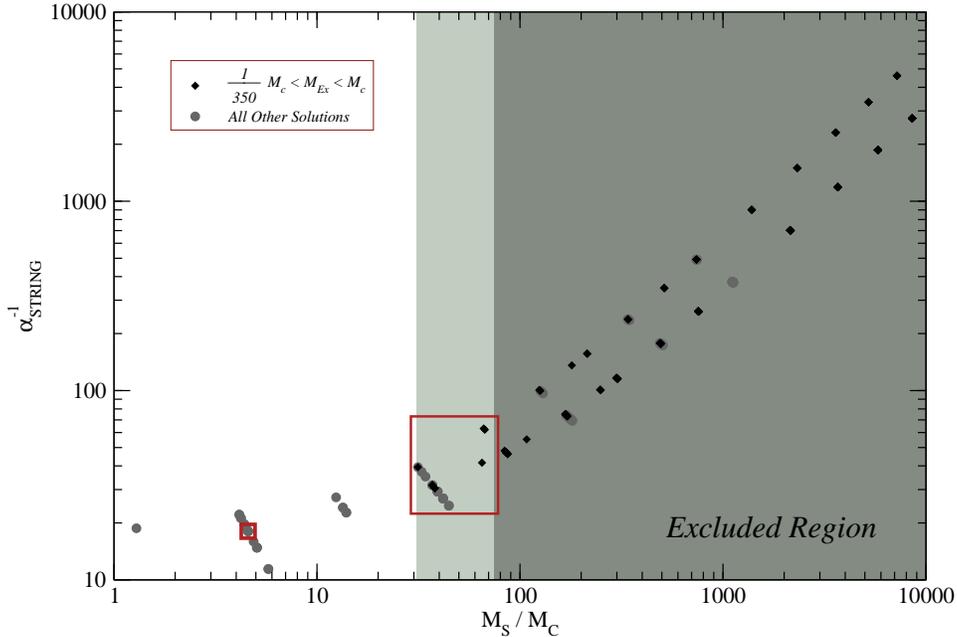}
        \caption{Here we show the correlation between the hierarchies in the problem.
        Quite generally, a small value of $\alphastring^{-1}$ requires a large hierarchy
        between the compactification scale and the exotic scale.  Again we show the
        excluded (darker green) and possibly testable (lighter green) models.
        The exact relationship between the ratio of $\mstring/\mc$ and the proton
        lifetime is given in Appendix \ref{sec:proton-decay}.  In particular, note
        the ``nice'' models (black diamonds) in the large red box, characterized by
        moderate hierarchies between all scales.  These models are collected in
        Table \ref{tab:interesting_models}.  Finally, note the one point in the
        small red box---this model is described in Section \ref{sec:top_down}.}
        \label{fig:scatter}
\end{figure}

In order to try and get a feel for the tunings involved in the above
conclusions, we can compare the GUT coupling constant (at the string
scale) with the ratio between the string scale and the
compactification scale.\footnote{The proton decay rate
$\Gamma(p\rightarrow \pi^0 e^+)$ is proportional to the fourth power
of the GUT coupling constant, see Appendix \ref{sec:proton-decay}.}
Further, we will separate the solutions based on the hierarchy
between the compactification scale and the exotic scale.  We plot
the result in Figure \ref{fig:scatter}. What we see is the
correlation between a long lived proton and a moderate hierarchy
between the compactification scale and the string scale, and between
the string scale and the Planck scale.  However, these moderate
hierarchies come at the cost of introducing a smaller and smaller
exotic scale, $\mex$.  This means that a long lived proton favors a
\textit{large} hierarchy between the compactification scale and the
exotic scale. The black diamonds represent those models with a
moderate ($<\mathcal{O}(350)$) hierarchy between the
compactification scale and the exotic scale. Most of these solutions
are already ruled out by proton decay constraints.  The gray shaded
circles represent those solutions for which there is a large
difference between the exotic scale and the compactification scale.

We would also like to point out the small set of solutions in the
large red box, for which there are only moderate hierarchies, and
which are consistent with the current bounds on dimension six
operators\footnote{See Appendix \ref{sec:proton-decay} for more
details.}.  Specifically, there seems to be a ``sweet spot'' where
all of the hierarchies in the problem are of $\mathcal{O}(100)$ or
so.  These models are highlighted in Table
\ref{tab:interesting_models}.  In particular, these models can all
be eliminated by improving the current bounds on proton decay from
dimension 6 operators by a factor of 50-100.

The fact that the data falls approximately on two straight lines is not
surprising, and is evidence of a power-law relationship between $\alphastring^{-1}$
and $\frac{\mstring}{\mc}$.  One can see this relationship as by
eliminating $\log\frac{\mstring}{\mex}$ between Equations (\ref{first_diff})
and (\ref{second_diff}).  We eventually find
\begin{equation}
   \log \alphastring^{-1} = A\log\frac{\mstring}{\mc} + B,
\end{equation}
where $A$ and $B$ are given in terms of the beta function coefficients and
$\log\frac{\mpl}{\mgut}$.  It is not surprising to find that the actual values
for $A$ and $B$ are roughly the same for all of the solutions, and that many
solutions give \textit{identical} values for $A$ and $B$.

\section{Unification, Decoupling of Exotics and Supersymmetry}
\label{sec:top_down}

Now that we understand what exotic matter we need to accommodate
unification, we can ask if an intermediate scale, $\mex$, is
consistent with decoupling of the other exotics.  The potential
difficulty can be summarized as follows: all of the (200,000+) mass
terms in the superpotential come from giving various MSSM singlets
VEVs.  Above, we have shown that unification depends on some exotics
receiving mass at the string scale, and some exotics receiving mass
at an intermediate scale.  This means that some singlets need to
have VEVs on the order of the string scale, $\mstring$, while other
singlets need to have VEVs on the order of the exotic mass scale,
$\mex$.  It is not obvious, \textit{a priori}, that we can do this
in a consistent way.  That is, decoupling with $D= F = 0$ was
checked in Reference \cite{Lebedev:2007hv}, but only for the case
where all of the singlet VEVs were of order the string scale.  In
light of gauge coupling unification, we are motivated to revisit the
previous conclusions.

As we will show, there is a very nice way to accommodate unification
in Model 1A, which relies only on moderate tunings.  The tunings
will be apparent when we address the question of $F = 0$ in Section
\ref{subsec:FFlat}.  In that section we will see how some numbers of
order the string scale must conspire to cancel some numbers of order
the exotic scale.

While Model 1A and Model 2 have similar sets of exotics, {\em they
have different super potentials}. So while it is possible to find
nice ways to accommodate unification within Model 1A, we find that
there \textit{does not} seem to be an easy way to assign singlet
VEVs in Model 2 such that we can accommodate unification.  This does
not mean that it is impossible to accommodate unification in Model
2, but it does make the process of assigning singlet VEVs an
exercise in fine tuning.

In what follows, we use the notation defined in Reference
\cite{Lebedev:2007hv} concerning the MSSM singlets.  In short, the
states labeled $s_i$ are singlets under the hidden sector and
visible sector gauge groups, while the states labeled $h_i$
transform as (hidden sector) \SU{2} doublets.  Some subset of the
$s_i$ and $h_i$ are expected to get non-zero VEVs, which defines a
vacuum configuration.  Again, we refer the reader to Reference
\cite{Lebedev:2007hv} for more details.

\subsection{Model 1A}

Let us first consider the issue of unification in Model 1A, where we
can solve the $F_i = 0$ equations exactly, giving us conditions on
the singlet VEVs to ensure that mass terms for the exotics do not
break supersymmetry at some high scale.  We must check that we can
consistently give some exotics intermediate scale mass, while
maintaining supersymmetry.

It turns out that giving only \textit{brane localized} exotic matter intermediate scale mass will not
give gauge coupling unification in this model.  This can be seen as follows: in order to get unification, we need
$b_3^{\textsc{ex}} - b_2^{\textsc{ex}} > 0$, otherwise the prediction for the string scale is
$\mstring \lesssim 10^{15} \gev$.  The states which contribute to this difference are (see Table \ref{tab:all_exotics},
for example)
\begin{equation}
   v \equiv (\mathbf{3},1)_{1/3,-2/3}, m\equiv(1,\mathbf{2})_{0,*} {\rm ~and~} y \equiv (1,\mathbf{2})_{0,0}.
\end{equation}
The mass matrices for the $y$ and $v$ turn out to be the same, which
means that we always get an equal number $v+\bar{v}$ and $y+y$ with
the same mass.  One can check in Table \ref{tab:proton_decay} that
there are no solutions in which the number of $v+\bar{v}$ is less
than or equal to the number of $y+y$.  Conversely, one can see this
from Equation (\ref{first_diff}).  If $n_2 \geq n_3$, the string
scale must be smaller than the GUT scale (assuming $\mstring >
\mex$), which (as we have previously argued) is not physical.  Thus
we \textit{must} give some bulk exotic matter intermediate scale
mass as well.

In giving bulk matter mass, we are severely limited in our options.
For one, the requirement that $b^{\mathcal{G}} < 0$ means that we
can only keep two extra pair of $\mathbf{6}+\mathbf{6}^c$ light.
Further, there is only one pair of extra down quarks and the states
$\delta + \bar{\delta}$.  In the first case, the extra $d+\bar{d}$
pair comes in an \SU6 multiplet with an extra $\ell + \bar{\ell}$,
both of which have $(++)$ boundary conditions, and both of which
couple in the same way to the singlet fields (to sixth order, and
likely to all orders). This means that they must get the same mass,
and we cannot get $b_3^{\textsc{ex}} - b_2^{\textsc{ex}} > 0$ in
this manner.  The remaining option is that we find an assignment of
singlet VEVs to give one pair of $\delta + \bar{\delta}$
intermediate scale mass.

Let us be a bit more explicit about how one would accomplish this, starting with a brief examination of the $\delta$s.
The mass matrix for the $\delta$s looks like
\begin{equation}
   \mathcal{M}_{\delta\bar{\delta}} = \left(\begin{array}{cccccc}
                    0&B_1&B_2&0&0&0\\
                    B_3&A_1&A_2&0&0&0\\
                    B_4&A_3&A_4&0&0&0\\
                    0&0&0&C_1&C_2&D_1\\
                    0&0&0&C_3&C_4&D_2\\
                    0&0&0&D_3&D_4&0\end{array}\right),
\end{equation}
where $A_i,B_i,C_i$ and $D_i$ are functions of singlet fields.  Let us concentrate on the upper left block of this matrix,
which involves only $A_i$ and $B_i$.  (The expressions for $C_i$ and $D_i$ are long and unenlightening, and not essential for the
discussion here.)   In general, the entries in the matrix have the following form:
\begin{eqnarray} \label{delta_mass_terms}
    A_i &\sim& \frac{1}{\mstring^5} s_1 \cdot s_5 \cdot s_6 \cdot s_{18} \cdot \left(h_1 \cdot h_{10} + h_2 \cdot h_9 \right), \\
    B_i &\sim& \frac{1}{\mstring^5} s_5 \cdot s_6 \left(h_{10}\cdot h_1 + h_9 \cdot h_2 \right) \cdot \left(h_1\cdot h_2 + s_{17}\cdot s_{18}\right).
\end{eqnarray}
Naively, diagonalizing this block gives one zero eigenvalue,
which means that there are two linear combinations of the $\delta$s
that are massless.  However, one must remember that the string
selection rules only give us the \textit{form} of the Yukawa
couplings, and not their exact magnitudes.  In general, this means
that we should be calculating $N$ point correlation functions on the
orbifold in order to get the \textit{exact} Yukawa couplings in the
theory.  In particular, it is important to remember that the
$\delta$s live at \textit{different} orbifold fixed points, and the
interaction eigenstates are a linear superposition of these
``orbifold eigenstates''.  Returning to Equation
(\ref{delta_mass_terms}), we see that if we require
\begin{eqnarray} \label{pre_VEV_assignment} \nonumber
   \langle s_1 \rangle &\sim& \mex, \\
   {\rm All~other~singlets~} &\sim& \mstring,
\end{eqnarray}
we naturally get one eigenstate with mass of order $\mex$, and five
heavy ($\sim \mstring$) eigenstates.  We note that there is some
dependence on $\langle s_1\rangle$ in the $C_i$ and $D_i$ at fifth
order in the singlets, however, there is \textit{no} dependence at
sixth order, suggesting that these terms (in general) dominate the
much smaller fifth order terms.

Next we consider the the $v+\bar{v}$ and $y+y$.  The mass matrices for these states are $2\times 2$ and identical,
and after they are diagonalized we find (ignoring constants of order one)
\begin{eqnarray} \nonumber \label{vev_cond}
m &\sim& s_{25}\Bigg\{1 + \frac{1}{\mstring^2} \left( s_{26} \cdot s_{15} + s_{26} \cdot s_{16}\right) + \frac{1}{\mstring^4} \left( s_{26}^2 \cdot s_{15} \cdot s_{16}  + s_{26}^2 \cdot s_{16}^2 + s_{26}^2 \cdot s_{15}^2 \right) + \\
&& + \frac{1}{\mstring^5} \left(s_{4} \cdot s_{6} \cdot s_{9} \cdot s_{30} \cdot s_{18}\right)\left(\frac{s_{11}\pm s_5}{s_{25}}\right)\Bigg\}.
\end{eqnarray}
It is clear that the following set of singlet VEVs is consistent with giving $2\times (v+\bar{v}) + 2\times (y+y)$
a mass at $\mex$:
\begin{eqnarray}  \nonumber \label{VEV_assignment}
   \langle s_1 \rangle \sim \langle s_{25}\rangle  &\sim& \mex, \\
   {\rm All~other~singlets~} &\sim& \mstring.
\end{eqnarray}
Note that we do rely here on some suppression in the sixth order term, so that it does not give an overwhelming
(i.e., $\mathcal{O}(\mstring)$) contribution to the mass term.  This may be viewed as an additional tuning in the singlet
VEVs, on the order of one part in ten or twenty.

Finally we check whether the VEV assignment (\ref{VEV_assignment})
is consistent with having some number of $(1,1)_{1,*} +
(1,1)_{-1,*}$ pairs with mass $\sim\mex$.  In general, the charged
singlet mass matrix (if we ignore the possibility of intermediate
scale mass for the $\bar{f}^+ + f^-$) is $14\times 14$ with equally
complicated eigenvalues, so we will omit the details of this
analysis.  Nevertheless, if we proceed in the same manner, we do
find two linear combinations of singlets ($s^+$ and $s^-$) whose
mass terms depend \textit{explicitly} on the VEV $\langle
s_{25}\rangle$, giving them naturally small mass terms.

We conclude that unification is possible \textit{in principle} in
Model 1A.  Specifically, in the absence of accidental cancellations,
and assuming that higher order terms in the superpotential are
negligible (such that the light linear combination of the $\delta$s
remains light), we have found one  version of Model 1A that gives us
gauge coupling unification.  Namely, if we assume order one
coefficients in the mass matrices, and that
\begin{equation}
  \langle s_{1}\rangle \sim \langle s_{25}\rangle \sim \mex, \qquad {\rm All~other~singlets~} \sim \mstring,
\end{equation}
we have exactly the following matter content in the theory with mass on the order of $\mex$:
\begin{eqnarray} \nonumber
   2 \times \left[v + \bar{v}\right] + 1 \times \left[y+y\right] + 2 \times \left[s^+ + s^-\right] + \left[\delta+\bar{\delta}\right].
\end{eqnarray}
This corresponds to the solution marked with an arrow ($\Rightarrow$) in Table \ref{tab:proton_decay_two} on page \pageref{tab:proton_decay_two}\footnote{Note
that the states $y$ are doublets under a hidden sector \SU{2}, so that $1 \times \left[y+y\right] \sim 2\times
\left[(1,\mbtwo)_{0,*}+(1,\mbtwo)_{0,*}\right]$}.  This gives us a prediction for the intermediate scale, the compactification
scale, the string scale, and proton decay coming from dimension six operators:
\begin{eqnarray} \nonumber
   \mex &\sim& 1.9 \times 10^{9} \gev , \\ \nonumber
   M_{\textsc{c}} &\sim& 2.2 \times 10^{17} \gev, \\ \nonumber
   \mstring &\sim& 1.0 \times 10^{18} \gev, \\
   \tau(p\rightarrow e^+ \pi^0) &\sim& 1.2 \times 10^{38} {\rm ~y}.
\end{eqnarray}
It is worth pointing out that this solution is not yet ruled out by
the current bounds on proton decay, a fact which was not guaranteed.
This model is pictured in the small red box in Figure
\ref{fig:scatter} on page \pageref{fig:scatter}.

We note that the other option that one may try is, for example
\begin{equation}
   \langle s_{11}\rangle+\langle s_5\rangle \sim \mstring, \qquad \langle s_{11}\rangle - \langle s_5\rangle \sim \mex, \qquad \langle s_{25} \rangle \sim \langle s_1 \rangle \sim \mex.
\end{equation}
This is a tuning to one part in $\mstring/\mex$, and is consistent
with $F=0$, which is discussed below.  This gives us one pair of
$v+\bar{v}$ and one $y$, and one pair of $\delta + \bar{\delta}$
with exotic scale mass, assuming that we \textit{can't} neglect the
sixth order term in Equation (\ref{vev_cond}).  The problem that one
may encounter is with the charged singlets. Taking $\langle s_{25}
\rangle \sim \mex$ generally gives one at least \textit{two} charged
singlets with mass at the intermediate scale, so one may need an
additional tuning in that sector of the theory in order to realize
one of the solutions in Table \ref{tab:proton_decay_two}.

\subsubsection{$\bs{F=0}$}
\label{subsec:FFlat}

Let us now comment on the compatibility of these solutions with the
constraint of $F=0$ in the case of Model 1A. If we set all of the
coefficients in the superpotential to one, the $F$ flatness
conditions can be solved exactly in this model.  In units where
$\mstring \equiv 1$, we find the following relationships among the
singlet VEVs:
\begin{eqnarray} \label{F_flatness}
   s_{22} &=& -\frac{1}{s_{20}+s_{21}} \left(h_1 h_2 + s_{17}s_{18}\right) - s_{23}, \\
   s_{26} &=& -\frac{1}{s_{15} + s_{16}}, \\ \nonumber \label{fine_tuned_VEV_relation_3}
   s_{1} &=& \frac{s_{15} + s_{16}}{s_{18}}\Big\{h_1 h_{10} + h_2 h_9 + s_{17} s_{25} + s_{18}s_{27}\\
   &&+ \left(s_{15} + s_{16}\right)s_{30} + \left(s_{20}+s_{21}\right) s_{31}\Big\}.
\end{eqnarray}
The task is to now assign arbitrary VEVs to everything
\textit{except} $s_{22}$, $s_{26}$, and $s_{1}$, and look for
solutions where $s_1 \sim s_{25} \sim \mex$.  The tuning in this
model is evident in Equation (\ref{fine_tuned_VEV_relation_3}). It
is clear that there must be a cancellation on the right hand side of
the equation to one part in $\mstring/\mex$. In general, one has no
trouble finding numerical solutions to these equations such that
$s_1 \sim s_{25} \sim \mex$, while all other singlets have VEVs near
the string scale.

One may object to the fact that we did not include superpotential
coefficients in Equation (\ref{F_flatness}) -
(\ref{fine_tuned_VEV_relation_3}), as it is clear that decoupling
depends on these coefficients \textit{not} being set to one. Solving
the $F$ flatness conditions with arbitrary superpotential
coefficients is a computationally intensive problem.  However, we
expect that the inclusion of such coefficients will not
significantly alter our conclusions.

\subsection{Model 2}
\label{sec:model-2-unification}

The exotic matter content of Model 2 is listed in Table \ref{tab:all_exotics}.  The \textit{brane localized} states which
contribute to the differential running $\alphathree^{-1} - \alphatwo^{-1}$ are
\begin{equation}
   v \equiv (\mathbf{3},1)_{1/3,-4/3}, m\equiv(1,\mathbf{2})_{0,*} {\rm ~and~} y \equiv (1,\mathbf{2})_{0,0}.
\end{equation}
In Model 2 we have
\begin{equation}
   4 \times (v + \bar{v}) + 2 \times (y+y) + 2\times (m + m) + 20\times (s^+ + s^-) + 2\times (x^+ + x^-),
\end{equation}
where $x^{\pm}$ are defined in Table \ref{tab:all_exotics}.

The mass matrix for the $v$ is a $4\times 4$ block diagonal matrix.  The blocks are both $2\times 2$, and
the upper block turns out to be equivalent to the ($2\times 2$) mass matrix for the $y$s.  By choosing
\begin{equation} \label{model_two_VEV_assignment}
   \langle h_2\rangle \sim \langle s_{43}\rangle \sim \mex, \qquad {\rm All~other~singlets~} \sim \mstring,
\end{equation}
we find $4\times (v+\bar{v}) + 2\times (y+y)$.  The problem with the VEV assignment in Equation (\ref{model_two_VEV_assignment})
is that we get too many charged singlets, so we will need to rely (heavily) on tuning arguments.
Thus we conclude that for Model 2 to be consistent with gauge coupling unification, we must arrange a
conspiracy among the singlet VEVs, such that we get intricate cancellations in the charged singlet sector.

\section{Conclusions}

We have addressed the question of gauge coupling unification in a
class of 15 ``mini-landscape'' models \cite{Lebedev:2007hv} with
properties very similar to the MSSM.    We analyze these $\E8 \times
\E8$ weakly coupled heterotic string models compactified on an
anisotropic orbifold with one large ($R$) and five small ($l_{\textsc{s}}$)
extra dimensions, where $R \gg l_{\textsc{s}}$ and $l_{\textsc{s}}$ is the string length.
All of these theories can then be described in terms of an effective
5D $\SU6$ orbifold GUT field theory with compactification scale $\mc
= 1/R$ and cut-off scale $\mstring = 1/l_{\textsc{s}}$.  $\SU6$ is broken to
the MSSM gauge group by orbifold boundary conditions at $\mc$ and
gauge couplings must unify at the cut-off scale, $\mstring$.
Moreover, in an orbifold GUT field theory, this is accomplished with
the aid of Kaluza-Klein modes which contribute to the RG running
above the compactification scale, $\mc$.

In all 15 models the electroweak Higgs doublets reside in the
(effective 4D, $N=2$) vector multiplet, hence the models satisfy
``gauge-Higgs unification." In addition the third family of quarks
and leptons are ``bulk" modes, while the two lighter families are
``brane" states. Although ``gauge-Higgs unification" may be
well-motivated by aesthetics, we prove in Appendix A that gauge
coupling unification is not possible if one only includes MSSM
states and their KK towers.   Thus it is necessary to also include
the possible contribution of vector-like exotics to the RG running.
To simplify the analysis, we assume a small set of exotics obtain
mass at a scale $\mex < \mc$ with the remainder obtaining mass at
$\mstring$.   Using an effective field theory analysis, we find many
solutions to gauge coupling unification labeled by the different
inequivalent sets of exotics with mass at $\mex$.    These solutions
are found in Tables \ref{tab:proton_decay} -
\ref{tab:interesting_models}, on pages \pageref{tab:proton_decay} -
\pageref{tab:interesting_models}.

We have analyzed two models in more detail (Models 1A and 2
\cite{Lebedev:2007hv}), since for these models we have the
superpotential up to order 6 in MSSM singlets.  In this case, we
have shown that one of our solutions (in Model 1A) is consistent
with string theory in a supersymmetric vacuum with $F = 0$,  if we
tune the singlet VEVs appropriately in Equation
(\ref{fine_tuned_VEV_relation_3}).   On the other hand, for the case
of Model 2, although there are many effective field theory
solutions, we have not been able to demonstrate the existence of a
simple string vacuum solution with $F = 0$.   In this case, a
solution may still be possible, however, it would require more
fine-tuning.

Since quarks and leptons of the first two families are located on an
effective $\SU5$ brane,  they are subject to proton decay processes
mediated by gauge exchange at the compactification scale $\mc$.
Moreover, since $\mc$ is generically less than the 4D GUT scale, the
proton decay rate for the process $p \rightarrow e^+ \pi^0$ is
enhanced.   Thus 80\% of the models satisfying gauge coupling
unification are excluded by Super-K bounds on proton decay.  Most of
the other models can be tested at a future proton decay detector.

All of the ``mini-landscape" models have an exact $R$ parity, so
they do not suffer from dimension 3 or 4 baryon and/or lepton number
violating processes.   Moreover, the LSP is stable and a possible
dark matter candidate.   However,  unlike 5D or 6D orbifold GUT
field theories studied in the literature, these models suffer from
uncontrolled dimension 5 operator contributions to proton decay.  In
particular, some of the vector-like exotics have quantum numbers of
color triplet Higgs multiplets.  When given mass at $\mstring$ or
$\mex$ they induce dimension 5 proton decay operators.   Although it
may be possible to fine-tune the coefficients of these operators to
be small, it would be preferable to have a symmetry argument.   This
problem needs to be addressed in any future string model building.

As noted, all of the models studied in this analysis have a 5D (or
6D) $\SU6$ orbifold GUT limit.  The complete spectrum of the 6D
model (prior to the final $\mathbb{Z}_2$ orbifold and Wilson line,
$A_2$) is given in Table \ref{tab:6dspectrum}.   It is very
interesting to note that the spectrum is identical with the spectrum
found in an $\E8 \times \E8$ heterotic string compactified on a
smooth $K_3 \times T^2$ manifold with instantons embedded in the
$\E8 \times \E8$ gauge groups \cite{Bershadsky:1996nh}.  This
suggests that these models may be obtained by the final
$\mathbb{Z}_2$ orbifolding of these smooth manifolds.

In conclusion, we have shown that gauge coupling unification may be
accommodated in the present class of string models.   However, a
simple solution, without including vector-like exotics below the
string scale, was not possible.   This appears to be a general
conclusion stemming from the particular implementation of
``gauge-Higgs unification" in these models.   Finally, any future
string model building needs to address the general problem of
uncontrolled dimension 5 baryon and lepton number violating
operators.

\section*{Acknowledgments}
We would like to thank Gerry Cleaver, Hyun-Min Lee and Hasan Y\"uksel for illuminating
conversations.  This work is supported under DOE grant number DOE/ER/01545-878.


\clearpage\newpage
\appendix

\section{Comparing two $\bs{\SU{6}}$ Orbifold GUTs}
\label{sec:very_similar}

The \SU{6} orbifold GUTs considered in this paper satisfy the
special property of gauge-Higgs unification.   This is also a
property of the 5D \SU6 orbifold GUT discussed in
Reference \cite{Hall:2001zb}.  It is instructive to compare this \SU{6}
model to one without gauge-Higgs unification, in particular the 5D
\SU{5} orbifold GUT discussed in Reference \cite{Hall:2001pg}.

In the models with gauge-Higgs unification, the Higgs multiplets
come from the 5D \textit{vector} multiplet ($V, \Phi$), both in the
adjoint representation of \SU{6}.   $V$ is the 4D gauge multiplet
and the 4D chiral multiplet $\Phi$ contains the Higgs doublets.
These states transform as follows under the orbifold parities
$(P\,\,P')$:
\begin{equation}
  V:\: \left( \begin{array}{ccc|cc|c}
    (+ +) & (+ +) & (+ +) & (+ -) & (+ -) & (- +) \\
    (+ +) & (+ +) & (+ +) & (+ -) & (+ -) & (- +) \\
    (+ +) & (+ +) & (+ +) & (+ -) & (+ -) & (- +) \\ \hline
    (+ -) & (+ -) & (+ -) & (+ +) & (+ +) & (- -) \\
    (+ -) & (+ -) & (+ -) & (+ +) & (+ +) & (- -) \\ \hline
    (- +) & (- +) & (- +) & (- -) & (- -) & (+ +)
  \end{array} \right)
\label{eq:V6trans}
\end{equation}
\begin{equation}
  \Phi:\: \left( \begin{array}{ccc|cc|c}
    (- -) & (- -) & (- -) & (- +) & (- +) & (+ -) \\
    (- -) & (- -) & (- -) & (- +) & (- +) & (+ -) \\
    (- -) & (- -) & (- -) & (- +) & (- +) & (+ -) \\ \hline
    (- +) & (- +) & (- +) & (- -) & (- -) & (+ +) \\
    (- +) & (- +) & (- +) & (- -) & (- -) & (+ +) \\ \hline
    (+ -) & (+ -) & (+ -) & (+ +) & (+ +) & (- -)
  \end{array} \right).
\label{eq:phi6trans}
\end{equation}
Note the appearance of the MSSM Higgs multiplets in $\Phi$ with $(+
+)$ boundary conditions, and it's partner in $V$ with $(--)$
boundary conditions.  These massive KK states contribute to a
logarithmic running of the gauge couplings with a term of the form
\begin{equation}
   \alpha_i^{-1} \supset -\frac{1}{4\pi}(b_i^{++} + b_i^{--})\log\frac{\mstring}{\mc}.
\end{equation}
We find for the model of Reference \cite{Hall:2001zb} (including just $V,
\Phi$ above)
\begin{equation} \label{HNS_bvalues}
  \vec{b}^{++} = (-9, -5, 3/5), \,\,\, \vec{b}^{--} = (3, -1, -9/5), \,\,\,\vec{b}^{++} + \vec{b}^{--} = (-6,-6,-6/5).
\end{equation}
(These numbers can be calculated using the values in Table
\ref{tab:all_b_values}.)   Again we stress that the only difference
between the models presented in this paper and that of Reference
\cite{Hall:2001zb} is that the third family lives in the bulk in our
constructions, which will only change these numbers by a universal
contribution. Indeed, one can check by comparing Equation
(\ref{HNS_bvalues}) with (\ref{b_values}) that the only difference
is a family universal contribution.

This can then be compared to an \SU{5} model without gauge-Higgs
unification \cite{Hall:2001pg}.   In this case the 5D gauge
multiplet includes the states, with their transformation under the
orbifold parities $(P\,\,P')$:

\begin{equation}
  V:\: \left( \begin{array}{ccc|cc}
    (+ +) & (+ +) & (+ +) & (+ -) & (+ -)  \\
    (+ +) & (+ +) & (+ +) & (+ -) & (+ -)  \\
    (+ +) & (+ +) & (+ +) & (+ -) & (+ -)  \\ \hline
    (+ -) & (+ -) & (+ -) & (+ +) & (+ +)  \\
    (+ -) & (+ -) & (+ -) & (+ +) & (+ +)
  \end{array} \right)
\label{eq:V5trans}
\end{equation}
\begin{equation}
  \Phi:\: \left( \begin{array}{ccc|cc}
    (- -) & (- -) & (- -) & (- +) & (- +)  \\
    (- -) & (- -) & (- -) & (- +) & (- +)  \\
    (- -) & (- -) & (- -) & (- +) & (- +)  \\ \hline
    (- +) & (- +) & (- +) & (- -) & (- -)  \\
    (- +) & (- +) & (- +) & (- -) & (- -)
  \end{array} \right).
\label{eq:phi5trans}
\end{equation}
The Higgs multiplets are contained in the chiral multiplets, $H_5 +
H_5^c$ and $H_{\bar 5} + H_{\bar 5}^c$, with parities
\begin{equation}
H_5, H_{\bar 5} :\:  \left( \begin{array}{c} (+ -) \\ (+ -) \\ (+ -)
\\ \hline (+ +) \\ (+ +) \end{array} \right).
\label{eq:higg5}
\end{equation}
\begin{equation}
H_5^c, H_{\bar 5}^c :\:  \left( \begin{array}{c} (- +) \\ (- +) \\
(- +)
\\ \hline (- -) \\ (- -) \end{array} \right).
\label{eq:higg5c}
\end{equation}
In this case, the $(- -)$ partners of the Higgs doublets appear in
chiral multiplets {\em not the gauge multiplet} as before.  Thus we
now find the beta function coefficients given by
\begin{equation} \label{HNS_bvalues_2}
  \vec{b}^{++} = (-9, -5, 3/5), \,\,\, \vec{b}^{--} = (3, 3, 3/5), \,\,\,\vec{b}^{++} + \vec{b}^{--} = (-6,-2, 6/5).
\end{equation}

To get relationships between the cutoff ($\mstring$) and the
compactification scale  ($\mc$), we can compare
$5\alphaone^{-1}(\mc)-3\alphatwo^{-1}(\mc)-2\alphathree^{-1}(\mc)$
and $\alphathree^{-1}(\mc)-\alphatwo^{-1}(\mc)$ in the orbifold GUT
and in the MSSM.  Including the threshold correction in Equation
(\ref{GQW_MSSM}) we find (for gauge-Higgs unification)
\begin{eqnarray} \nonumber \label{HNS_diff_relations_su6}
\log\frac{\mgut}{\mc} &=& \frac{2}{3}\log\frac{\mstring}{\mc} + \frac{1}{3}, \\
\log\frac{\mstring}{\mgut} &=& -\frac{3}{2}
\end{eqnarray}
The factors of $\frac{1}{3}$ and $-\frac{3}{2}$ come from the
threshold  correction applied at $\mgut$.  These equations
implicitly assume the relation $\mc \leq \mgut, \mstring$, however,
the solution to the equation gives the unphysical relation $\mc >
\mgut > \mstring$.   This is the main reason we need to rely on
``light" exotics.   On the other hand,  for the \SU{5} orbifold GUT
we find
\begin{eqnarray} \nonumber \label{HNS_diff_relations_su5}
\log\frac{\mgut}{\mc} &=& \frac{2}{3}\log\frac{\mstring}{\mc} + \frac{1}{3}, \\
\log\frac{\mgut}{\mc} &=& \frac{1}{2}\log\frac{\mstring}{\mc} +
\frac{3}{2}
\end{eqnarray}
which gives the physically acceptable solution
$\log\frac{\mstring}{\mgut} = 2$ and $\log\frac{\mgut}{\mc} = 5$. We
thus conclude that simple gauge-Higgs unification in 5D \SU{6} is
not viable.

In Reference \cite{Hall:2001zb} an $N=2$ model with gauge-Higgs
unification in 6D (or $N=4$ in 4D) was also considered.  In this
case the Higgs multiplet and its $(- -)$ partners are contained in
chiral adjoints.    Gauge coupling unification works in this model.
Unfortunately, we do not know how to obtain such a model from the
heterotic string.

Of course, the additional problem concerning gauge coupling
unification in the context of the heterotic string is the need to
match the low energy values of the coupling constants given values
of $\mc$ and $\mstring$. In particular, we must satisfy the relation
\begin{equation}
\alphastring^{-1} = \frac{1}{8}\left(\frac{\mpl}{\mstring}\right)^2.
\end{equation}
In most cases, with $\mc \leq \mgut < \mstring$, the power law
running due to the KK modes is required, i.e.
\begin{equation}
\alpha_{i}^{-1}(\mc) \supset \alphastring^{-1} +
\frac{b^{\mathcal{G}}}{2\pi}\left(\frac{\mstring}{\mc} - 1\right) +
{\rm Log \;\; terms} \sim \mathcal{O}(10).
\end{equation}

\section{Constraints from Proton Decay}
\label{sec:proton-decay}

\subsection{Dimension 6 Operators}

The gauge bosons in GUTs can mediate proton decay via effective
dimension 6 operators. The best bounds on proton decay come from the
channel $p\rightarrow e^+ + \pi^0$, and current (published) experimental limits
are\cite{Yao:2006px} \beqn \tau(p\rightarrow e^+ + \pi^0) > 1.6
\times 10^{33} \rm{~yr}. \eeqn In this paper, we are looking at an
\SU6 GUT in five dimensions, which is broken to either \SU5 or
$\SU4\times \SU2$ on the branes.   The dangerous operators come from
\SU5 gauge boson ($\mathbf{X}$) exchange and have been calculated in
Reference \cite{Hisano:2000dg}.  In a 4-d \SU5 GUT, the effective
lagrangian leading to proton decay from $\mathbf{X}$ boson exchange
is given by
\begin{equation}
\mathcal{L}_{\mathrm{eff}} = \frac{g_{\gut}^2}{2 M_{\mathbf{X}}^2}  J^{\mu}J_{\mu}^*,
\end{equation}
where
\begin{equation}
J^{\mu} = -(l)^*\bar{\sigma}^{\mu}d^c + (u^c)^*\bar{\sigma}^{\mu}q + (q)^* \bar{\sigma}^{\mu} e^c + \mathrm{h.c.}
\end{equation}
The operators which lead to proton decay are given by
\begin{equation}
\mathcal{L}_{\mathrm{eff}} = -\frac{g_{\gut}^2}{2 M_{\mathbf{X}}^2} \sum_{i,j}\left[(q^*_i \bar{\sigma}^{\mu}u^c_i)(\ell^*\bar{\sigma}_{\mu}d^c_j) + (q^*_i \bar{\sigma}^{\mu}e^c_i)  (q_j^*\bar{\sigma}_{\mu}u^c_j)\right].
\end{equation}
The decay rate of $p\rightarrow \pi^0 e^+$ in the 4-d theory is given by
\begin{equation} \label{decay_rate}
\Gamma(p\rightarrow \pi^0 e^+) = \frac{\left(m_p^2-m_{\pi}^2\right)^2}{64\pi m_p^3 f_{\pi}^2} \beta_{\textsc{lat}}^2 A^2 \frac{g_{\gut}^4}{M_{\mathbf{X}}^4} \left(1+D+F\right)^2 \left[\left(1+|V_{ud}|^2\right)^2 + 1\right].
\end{equation}
These formulae will receive modifications in our model,
based on the fact that there is a relationship between
the string scale, the Planck scale and the coupling constant (see
Equation (\ref{het_string_BC}) ), and that the whole tower of
KK modes associated with the \SU5 gauge bosons will contribute
to the decay rate.

Explicitly, the decay rate goes like $g_{\gut}^4$.  We replace this by
\begin{equation}
g_{\gut}^4 \rightarrow (4\pi)^2\alphastring^2 = 64\times(4\pi)^2\times\left(\frac{\mstring}{\mpl}\right)^4.
\end{equation}
Next, we should consider the relationship between the
compactification scale and the $\mathbf{X}$ boson mass.
The \SU5 gauge bosons have $(+-)$ boundary conditions,
and masses of $m_n = \left(n+\frac{1}{2}\right) \mc$.
Proton decay can proceed by exchange of any of the
tower of KK modes, which suggests we take
\begin{equation}
\frac{1}{M_{\mathbf{X}}^2} \rightarrow 2\times\frac{1}{\mc^2} \sum_{n=0}^{\infty} \frac{1}{\left(n+\frac{1}{2}\right)^2} = \frac{\pi^2}{\mc^2}.
\end{equation}
The factor of two comes from the fact that the KK modes of
the gauge bosons are normalized differently than the zero
modes \cite{Hall:2001pg}.\footnote{Equivalently, one can understand
this factor as the Kaluza-Klein tower of gauge
bosons coupling \textit{more strongly} to the fermions by a
factor of $\sqrt{2}$, which corresponds to rescaling
$g_{\gut}\rightarrow \sqrt{2}g_{\gut}$.}  Including all
corrections, we make the replacement
\begin{equation}
\frac{g_{\gut}^4}{M_{\mathbf{X}}^4} \rightarrow 64 \times (4\pi)^2 \times \left(\frac{\mstring}{\mpl}\right)^4 \times \frac{\pi^4}{\mc^4}
\end{equation}

In our 5-d orbifold GUT, we find
\begin{equation}
\Gamma(p\rightarrow \pi^0 e^+) \cong 4.00 \times 10^{-73} \left(\frac{\mstring}{\mc}\right)^4 \gev.
\end{equation}
where we have used $A=3.4,D=0.80$ and $F=0.44$, and
$\beta_{\textsc{lat}} \simeq 0.011 \gev^3$ \cite{Aoki:2006ib}.
For the proton lifetime, we find
\begin{equation}
\tau(p\rightarrow \pi^0 e^+) \cong 5.21 \times 10^{40} \left(\frac{\mc}{\mstring}\right)^4 \mathrm{yr}.
\end{equation}
This corresponds to an upper limit on the ratio between the string scale and the compactification scale of
\begin{equation}
\frac{\mstring}{\mc} \lesssim 75.
\end{equation}
Alternatively, given a (typical) string scale of about $5\times 10^{17} \gev$, this corresponds to
\begin{equation}
\mc \gtrsim 6.6 \times 10^{15} \gev.
\end{equation}
An interesting difference between this result and the result one
typically finds in an orbifold GUT (see for example Reference
\cite{Alciati:2005ur}) is that the proton lifetime no longer
scales like the compactification scale directly, but as a ratio
of scales.  This means that the compactification scale can
be smaller than $\mc \sim 6.6 \times 10^{15} \gev$
if the string scale is sufficiently small, which means that the
underlying GUT is very weakly coupled ($\alpha_{\gut} << 1$).\footnote{This
may correspond to a region where the string coupling constant
$g_{\textsc{string}} \sim e^{\phi}$ (where $\phi$ is the
dilaton field) is no longer small.  This is undesirable,
as we wish to embed these models in the weakly coupled
heterotic string \cite{inprogress}.}
We note that this is an additional constraint that has no analogy in typical
orbifold GUT model building, imposed by the relationship
between the coupling constant, Newton's constant, and $\alpha'$.
Finally, in the interesting limit that $\mc\rightarrow\mstring$,
we find the upper bound on proton lifetime in this class of
models: $\tau(p\rightarrow\pi^0e^+) \lesssim 5.21 \times 10^{40}$ yr.

\subsection{Dimension 5 Operators}

In supersymmetric theories, the proton may decay via dimension five
operators as well. In  the mini-landscape models
\cite{Lebedev:2007hv}, the
$(\mathbf{3},1)_{-2/3,-2/3}+(\overline{\mathbf{3}},1)_{2/3,2/3}$
states, called $\delta$ and $\overline{\delta}$, can mediate proton
decay via dimension five operators---they have the same gauge
quantum numbers as color triplet Higgses.  It was shown in Reference
\cite{Dermisek:2000hr} that the effective mass of the color triplet
Higgsino $M_{\tilde H} \sim 10^{18}-10^{21} \gev$ has to be much
larger than the (four dimensional) GUT scale in order to evade
bounds on $p\rightarrow K^+ \bar{\nu}$, depending on the soft SUSY
breaking parameters.

The $\delta$ particles have the same quantum numbers as color
triplet Higgses, thus we expect similar bounds for them (assuming
they couple to quarks and leptons with small effective Yukawa
couplings). Unfortunately, to make matters worse, it was found in
Reference \cite{Lebedev:2007hv} that the $\delta$ states have tree
level coupling to the quarks in the superpotential, and so the
coupling is naturally of order one, i.e.~\textit{not} suppressed by
Yukawa factors as they are in the typical dimension five proton
decay operator.  However, by carefully adjusting the singlet VEVs
that describe the $\delta$, $\bar{\delta}$, interactions, this
problem can be avoided, but currently, we are lacking a mechanism
that would naturally suppress this decay channel for the proton.

\newpage

\section{\label{app:model_two} Miscellany}

\begin{table}[h!]
\caption{Spectrum of Model 1 of the mini-landscape search \cite{Lebedev:2007hv}. From the viewpoint of the 5-dimensional theory, all states that are not localize in the $\SO{4}$ torus ($U$, $T_2$, $T_4$) are bulk modes. The symbols $\bullet$, $\bigstar$, $\blacksquare$, $\blacktriangle$ indicate the localization of the brane modes in the $\SO{4}$ torus, compare Figure \ref{fig:5d_picture} on page \pageref{fig:5d_picture}.}
\label{tab:spectrum_model1}
\centering
\normalsize
\renewcommand{\arraystretch}{1.4}
\scalebox{0.75}{
\begin{tabular}{|l|c||l|c||l|c|l|c|}
\hline
\multicolumn{2}{|c||}{U}  &                                                                 \multicolumn{2}{|c||}{$T_3$}  &                                                       \multicolumn{4}{|c|}{$T_5$} \\
\hline
$1 \times  ( \bs{3} , \bs{2} )_{\ptm1/3, \ptm1/3 }$         & bulk  &  $4 \times  ( \bs{1} , \bs{1} )_{\ptm1, \ptm3 }$      & $\blacktriangle$  &
    $1 \times  ( \bs{3} , \bs{2} )_{\ptm1/3, \ptm1/3 }$         & $\bigstar$        & $1 \times  ( \bs{1} , \bs{1} )_{\text{-}1, \text{-}3 }$   & $\blacksquare$  \\
$1 \times  ( \bsb{3}, \bs{1} )_{\text{-}4/3, \text{-}1/3 }$     & bulk  &  $4 \times  ( \bs{1} , \bs{1} )_{\text{-}1, \text{-}3 }$  & $\blacktriangle$  &
    $1 \times  ( \bs{3} , \bs{2} )_{\ptm1/3, \ptm1/3 }$         & $\bullet$         & $1 \times  ( \bs{1} , \bs{1} )_{\ptm1, \text{-}3 }$       & $\blacktriangle$  \\
$1 \times  ( \bs{1} , \bs{2} )_{\ptm1, \ptm0 }$         & bulk  &  $4 \times  ( \bs{1} , \bs{1} )_{\ptm1, \ptm3 }$      & $\blacksquare$    &
    $1 \times  ( \bsb{3}, \bs{1} )_{\text{-}4/3, \text{-}1/3 }$     & $\bigstar$        & $1 \times  ( \bs{1} , \bs{1} )_{\ptm1, \ptm3 }$       & $\blacktriangle$  \\
$1 \times  ( \bs{1} , \bs{2} )_{\text{-}1, \ptm0 }$         & bulk  &  $4 \times  ( \bs{1} , \bs{1} )_{\text{-}1, \text{-}3 }$  & $\blacksquare$    &
    $1 \times  ( \bsb{3}, \bs{1} )_{\text{-}4/3, \text{-}1/3 }$     & $\bullet$         & $1 \times  ( \bs{1} , \bs{1} )_{\ptm1, \ptm3 }$       & $\blacksquare$  \\
$1 \times  ( \bs{1} , \bs{1} )_{\ptm2, \ptm1 }$         & bulk  &  $2 \times  ( \bs{1} , \bs{1} )_{\ptm1, \ptm2 }$      & $\blacktriangle$  &
    $1 \times  ( \bsb{3}, \bs{1} )_{\ptm2/3, \text{-}1/3 }$     & $\bigstar$        & $1 \times  ( \bs{1} , \bs{1} )_{\ptm1, \text{-}3 }$       & $\blacksquare$  \\
$4 \times  ( \bs{1} , \bs{1} )_{\ptm0, \text{-}1 }$         & bulk  &  $2 \times  ( \bs{1} , \bs{1} )_{\text{-}1, \text{-}2 }$  & $\blacktriangle$  &
    $1 \times  ( \bsb{3}, \bs{1} )_{\ptm2/3, \text{-}1/3 }$     & $\bullet$         & $2 \times  ( \bs{1} , \bs{1} )_{\ptm1, \text{-}2 }$       & $\blacktriangle$  \\
$5 \times  ( \bs{1} , \bs{1} )_{\ptm0, \ptm1 }$         & bulk  &  $2 \times  ( \bs{1} , \bs{1} )_{\ptm1, \ptm2 }$      & $\blacksquare$    &
    $1 \times  ( \bsb{3}, \bs{1} )_{\text{-}1/3, \ptm8/3 }$     & $\blacktriangle$      & $2 \times  ( \bs{1} , \bs{1} )_{\text{-}1, \ptm2 }$       & $\blacktriangle$  \\
$2 \times  ( \bs{1} , \bs{1} )_{\ptm0, \ptm0 }$         & bulk  &  $2 \times  ( \bs{1} , \bs{1} )_{\text{-}1, \text{-}2 }$  & $\blacksquare$    &
    $1 \times  ( \bs{3} , \bs{1} )_{\ptm1/3, \text{-}8/3 }$     & $\blacktriangle$      & $2 \times  ( \bs{1} , \bs{1} )_{\ptm1, \text{-}2 }$       & $\blacksquare$  \\
\cline{1-2}
\multicolumn{2}{|c||}{$T_2$}                        &  $1 \times  ( \bs{1} , \bs{1} )_{\ptm1, \ptm2 }$      & $\blacktriangle$  &
    $1 \times  ( \bsb{3}, \bs{1} )_{\text{-}1/3, \ptm8/3 }$     & $\blacksquare$    & $2 \times  ( \bs{1} , \bs{1} )_{\text{-}1, \ptm2 }$       & $\blacksquare$  \\
\cline{1-2}
$3 \times  ( \bsb{3} , \bs{1} )_{\ptm2/3, \ptm2/3 }$        & bulk  & $1 \times  ( \bs{1} , \bs{1} )_{\ptm1, \text{-}2 }$       & $\blacktriangle$  &
    $1 \times  ( \bs{3} , \bs{1} )_{\ptm1/3, \text{-}8/3 }$     & $\blacksquare$    & $1 \times  ( \bs{1} , \bs{1} )_{\ptm0, \ptm5 }$       & $\bigstar$  \\
$3 \times  ( \bs{3} , \bs{1} )_{\text{-}2/3, \text{-}2/3 }$     & bulk  & $1 \times  ( \bs{1} , \bs{1} )_{\ptm1, \text{-}2 }$       & $\blacksquare$    &
    $1 \times  ( \bs{1} , \bs{2} )_{\text{-}1, \text{-}1 }$     & $\bigstar$        & $1 \times  ( \bs{1} , \bs{1} )_{\ptm0, \text{-}5 }$       & $\bigstar$  \\
$2 \times  ( \bsb{3} , \bs{1} )_{\ptm 2/3, \text{-}1/3 }$   & bulk  &       $1 \times  ( \bs{1} , \bs{1} )_{\text{-}1, \ptm2 }$     & $\blacksquare$    &
    $1 \times  ( \bs{1} , \bs{2} )_{\text{-}1, \text{-}1 }$     & $\bullet$         & $1 \times  ( \bs{1} , \bs{1} )_{\ptm0, \ptm5 }$       & $\bullet$  \\
$1 \times  ( \bs{1} , \bs{2} )_{\ptm1, \ptm1 }$         & bulk  & $1 \times  ( \bs{1} , \bs{1} )_{\ptm0, \ptm6 }$       & $\bigstar$        &
    $1 \times  ( \bs{1} , \bs{2} )_{\ptm0, \text{-}3 }$         & $\blacktriangle$      & $1 \times  ( \bs{1} , \bs{1} )_{\ptm0, \text{-}5 }$       & $\bullet$  \\
$3 \times  ( \bs{1} , \bs{1} )_{\ptm0, \ptm5 }$         & bulk  & $1 \times  ( \bs{1} , \bs{1} )_{\ptm0, \text{-}6 }$       & $\bigstar$        &
    $1 \times  ( \bs{1} , \bs{2} )_{\ptm0, \ptm3 }$         & $\blacktriangle$      & $2 \times  ( \bs{1} , \bs{1} )_{\ptm0, \ptm3 }$       & $\bigstar$ \\
$6 \times  ( \bs{1} , \bs{1} )_{\ptm0, \ptm3 }$         & bulk  & $1 \times  ( \bs{1} , \bs{1} )_{\ptm0, \ptm6 }$       & $\bullet$         &
    $1 \times  ( \bs{1} , \bs{2} )_{\ptm0, \ptm3 }$         & $\blacksquare$    & $2 \times  ( \bs{1} , \bs{1} )_{\ptm0, \text{-}3 }$       & $\bigstar$ \\
$4 \times  ( \bs{1} , \bs{1} )_{\ptm0, \ptm2 }$         & bulk  & $1 \times  ( \bs{1} , \bs{1} )_{\ptm0, \text{-}6 }$       & $\bullet$         &
    $1 \times  ( \bs{1} , \bs{2} )_{\ptm0, \text{-}3 }$         & $\blacksquare$    & $2 \times  ( \bs{1} , \bs{1} )_{\ptm0, \ptm3 }$       & $\bullet$  \\
$4 \times  ( \bs{1} , \bs{1} )_{\ptm0, \text{-}2 }$         & bulk  & $2 \times  ( \bs{1} , \bs{1} )_{\ptm0, \text{-}2 }$       & $\bigstar$        &
    $1 \times  ( \bs{1} , \bs{2} )_{\ptm0, \ptm2 }$         & $\blacktriangle$      & $2 \times  ( \bs{1} , \bs{1} )_{\ptm0, \text{-}3 }$       & $\bullet$  \\
$5 \times  ( \bs{1} , \bs{1} )_{\ptm0, \ptm1 }$         & bulk  & $2 \times  ( \bs{1} , \bs{1} )_{\ptm0, \ptm2 }$       & $\bigstar$        &
    $1 \times  ( \bs{1} , \bs{2} )_{\ptm0, \text{-}2 }$         & $\blacktriangle$      & $1 \times  ( \bs{1} , \bs{1} )_{\ptm0, \text{-}1 }$       & $\bigstar$ \\
$2 \times ( \bs{1} , \bs{1} )_{\ptm0, \text{-}1 }$      & bulk  & $2 \times  ( \bs{1} , \bs{1} )_{\ptm0, \text{-}2 }$       & $\bullet$         &
    $1 \times  ( \bs{1} , \bs{2} )_{\ptm0, \ptm2 }$         & $\blacksquare$    & $1 \times  ( \bs{1} , \bs{1} )_{\ptm0, \text{-}1 }$       & $\bullet$ \\
$21 \times ( \bs{1} , \bs{1} )_{\ptm0, \ptm0 }$         & bulk  & $2 \times  ( \bs{1} , \bs{1} )_{\ptm0, \ptm2 }$       & $\bullet$         &
    $1 \times  ( \bs{1} , \bs{2} )_{\ptm0, \text{-}2 }$         & $\blacksquare$    & $1 \times  ( \bs{1} , \bs{1} )_{\ptm0, \ptm1 }$       & $\bigstar$  \\
\cline{1-4}
\multicolumn{2}{|c||}{$T_4$}                        &                                               &               &                                   $2 \times  ( \bs{1} , \bs{2} )_{\ptm0, \ptm0 }$         & $\blacktriangle$      & $1 \times  ( \bs{1} , \bs{1} )_{\ptm0, \ptm1 }$       & $\bigstar$  \\
\cline{1-2}
$3 \times  ( \bsb{3} , \bs{1} )_{\ptm2/3, \ptm2/3 }$        & bulk  &                               &           &                                   $2 \times  ( \bs{1} , \bs{2} )_{\ptm0, \ptm0 }$         & $\blacksquare$    & $1 \times  ( \bs{1} , \bs{1} )_{\ptm0, \ptm1 }$       & $\bullet$ \\
$3 \times  ( \bs{3} , \bs{1} )_{\text{-}2/3, \text{-}2/3 }$     & bulk  &                               &           &                                   $1 \times  ( \bs{1} , \bs{1} )_{\ptm2, \ptm1 }$         & $\bigstar$        & $1 \times  ( \bs{1} , \bs{1} )_{\ptm0, \ptm1 }$       & $\bullet$ \\
$1 \times  ( \bsb{3} , \bs{1} )_{\text{-} 2/3, \ptm1/3 }$   & bulk  &                               &           &                                   $1 \times  ( \bs{1} , \bs{1} )_{\ptm2, \ptm1 }$         & $\bullet$         & $8 \times  ( \bs{1} , \bs{1} )_{\ptm0, \ptm0 }$       & $\bigstar$  \\
$2 \times  ( \bs{1} , \bs{2} )_{\text{-}1, \text{-}1 }$     & bulk  &                               &           &                                   $1 \times  ( \bs{1} , \bs{1} )_{\text{-}1, \text{-}3 }$     & $\blacktriangle$  & $8 \times  ( \bs{1} , \bs{1} )_{\ptm0, \ptm0 }$       & $\bigstar$  \\
$3 \times  ( \bs{1} , \bs{1} )_{\ptm0, \text{-}5 }$         & bulk  &                               &           &                                   $1 \times  ( \bs{1} , \bs{1} )_{\text{-}1, \ptm3 }$     & $\blacktriangle$  & $6 \times ( \bs{1} , \bs{1} )_{\ptm0, \ptm0 }$        & $\bullet$ \\
$6 \times  ( \bs{1} , \bs{1} )_{\ptm0, \text{-}3 }$         & bulk  &                               &           &                                   $1 \times  ( \bs{1} , \bs{1} )_{\text{-}1, \ptm3 }$         & $\blacksquare$    & $6 \times ( \bs{1} , \bs{1} )_{\ptm0, \ptm0 }$        & $\bullet$  \\
\cline{5-8}
$2 \times  ( \bs{1} , \bs{1} )_{\ptm0, \text{-}2 }$         & bulk  & & & & & &   \\
$2 \times  ( \bs{1} , \bs{1} )_{\ptm0, \ptm2 }$         & bulk  & & & & & &   \\
$1 \times  ( \bs{1} , \bs{1} )_{\ptm0, \ptm1 }$         & bulk  & & & & & &   \\
$4 \times ( \bs{1} , \bs{1} )_{\ptm0, \text{-}1 }$      & bulk  & & & & & &   \\
$21 \times ( \bs{1} , \bs{1} )_{\ptm0, \ptm0 }$         & bulk  & & & & & &   \\
\hline
\end{tabular}
}
\end{table}

\begin{table}[h!]
\caption{Spectrum of Model 2 of the mini-landscape search \cite{Lebedev:2007hv}. From
the viewpoint of the 5-dimensional theory, all states that are not localize in the
$\SO{4}$ torus ($U$, $T_2$, $T_4$) are bulk modes. The symbols $\bullet$, $\bigstar$,
$\blacksquare$, $\blacktriangle$ indicate the localization of the brane modes in the
$\SO{4}$ torus, compare Figure \ref{fig:5d_picture} on page \pageref{fig:5d_picture}.}
\label{tab:spectrum_model2}
\centering
\normalsize
\renewcommand{\arraystretch}{1.4}
\scalebox{0.75}{
\begin{tabular}{|l|c||l|c||l|c|l|c|}
\hline
\multicolumn{2}{|c||}{U}  &                                                                 \multicolumn{2}{|c||}{$T_3$}  &                                                       \multicolumn{4}{|c|}{$T_5$} \\
\hline
$1 \times  ( \bs{3} , \bs{2} )_{\phantom{\text{-}}1/3, \phantom{\text{-}}1/3 }$ & bulk  &  $1 \times  ( \bs{3} , \bs{1} )_{\phantom{\text{-}}1/3, \text{-}8/3 }$ & $\blacktriangle$   &       $1 \times  ( \bs{3} , \bs{2} )_{\phantom{\text{-}}1/3, \phantom{\text{-}}1/3 }$ & $\bigstar$  &  $2 \times  ( \bs{1} , \bs{1} )_{\text{-}1, \text{-}2 }$ & $\blacksquare$  \\
$1 \times  ( \bsb{3}, \bs{1} )_{\text{-}4/3, \text{-}1/3 }$ & bulk   &             $1 \times  ( \bsb{3}, \bs{1} )_{\text{-}1/3, \phantom{\text{-}}8/3 }$ & $\blacktriangle$   &       $1 \times  ( \bs{3} , \bs{2} )_{\phantom{\text{-}}1/3, \phantom{\text{-}}1/3 }$ & $\bullet$   &  $2 \times  ( \bs{1} , \bs{1} )_{\phantom{\text{-}}1, \phantom{\text{-}}1 }$ & $\blacktriangle$  \\
$1 \times  ( \bs{1} , \bs{2} )_{\phantom{\text{-}}1, \phantom{\text{-}}0 }$ & bulk   &     $1 \times  ( \bs{3} , \bs{1} )_{\phantom{\text{-}}1/3, \text{-}8/3 }$ & $\blacksquare$   &       $1 \times  ( \bsb{3}, \bs{1} )_{\text{-}4/3, \text{-}1/3 }$ & $\bigstar$  &               $2 \times  ( \bs{1} , \bs{1} )_{\phantom{\text{-}}1, \text{-}1 }$ & $\blacktriangle$  \\
$1 \times  ( \bs{1} , \bs{2} )_{\text{-}1, \phantom{\text{-}}0 }$ & bulk   &           $1 \times  ( \bsb{3}, \bs{1} )_{\text{-}1/3, \phantom{\text{-}}8/3 }$ & $\blacksquare$   &       $1 \times  ( \bsb{3}, \bs{1} )_{\text{-}4/3, \text{-}1/3 }$ & $\bullet$   &               $2 \times  ( \bs{1} , \bs{1} )_{\text{-}1, \phantom{\text{-}}1 }$ & $\blacktriangle$  \\
$1 \times  ( \bs{1} , \bs{1} )_{\phantom{\text{-}}2, \phantom{\text{-}}1 }$ & bulk   &     $3 \times  ( \bs{1} , \bs{1} )_{\phantom{\text{-}}1, \phantom{\text{-}}3 }$ & $\blacktriangle$   & $1 \times  ( \bsb{3}, \bs{1} )_{\phantom{\text{-}}2/3, \text{-}1/3 }$ & $\bigstar$  &           $2 \times  ( \bs{1} , \bs{1} )_{\text{-}1, \text{-}1 }$ & $\blacktriangle$  \\
$1 \times  ( \bs{1} , \bs{1} )_{\phantom{\text{-}}0, \text{-}2 }$ & bulk   &           $3 \times  ( \bs{1} , \bs{1} )_{\text{-}1, \text{-}3 }$ & $\blacktriangle$   &                 $1 \times  ( \bsb{3}, \bs{1} )_{\phantom{\text{-}}2/3, \text{-}1/3 }$ & $\bullet$   &           $2 \times  ( \bs{1} , \bs{1} )_{\phantom{\text{-}}1, \text{-}1 }$ & $\blacksquare$  \\
$1 \times  ( \bs{1} , \bs{1} )_{\phantom{\text{-}}0, \phantom{\text{-}}2 }$ & bulk   &     $3 \times  ( \bs{1} , \bs{1} )_{\phantom{\text{-}}1, \phantom{\text{-}}3 }$ & $\blacksquare$   & $1 \times  ( \bsb{3}, \bs{1} )_{\text{-}1/3, \phantom{\text{-}}5/3 }$ & $\blacktriangle$   &          $2 \times  ( \bs{1} , \bs{1} )_{\phantom{\text{-}}1, \phantom{\text{-}}1 }$ & $\blacksquare$  \\
$8 \times  ( \bs{1} , \bs{1} )_{\phantom{\text{-}}0, \phantom{\text{-}}1/2 }$ & bulk  &    $3 \times  ( \bs{1} , \bs{1} )_{\text{-}1, \text{-}3 }$ & $\blacksquare$   &                   $1 \times  ( \bs{3} , \bs{1} )_{\phantom{\text{-}}1/3, \text{-}5/3 }$ & $\blacktriangle$   &            $2 \times  ( \bs{1} , \bs{1} )_{\text{-}1, \phantom{\text{-}}1 }$ & $\blacksquare$  \\
$1 \times  ( \bs{1} , \bs{1} )_{\phantom{\text{-}}0, \phantom{\text{-}}0 }$ & bulk   &     $1 \times  ( \bs{1} , \bs{1} )_{\phantom{\text{-}}1, \text{-}2 }$ & $\blacktriangle$   &       $1 \times  ( \bsb{3}, \bs{1} )_{\text{-}1/3, \phantom{\text{-}}5/3 }$ & $\blacksquare$   &          $2 \times  ( \bs{1} , \bs{1} )_{\text{-}1, \text{-}1 }$ & $\blacksquare$  \\
\cline{1-2}
\multicolumn{2}{|c||}{$T_2$}  &                                    $1 \times  ( \bs{1} , \bs{1} )_{\text{-}1, \phantom{\text{-}}2 }$ & $\blacktriangle$   &       $1 \times  ( \bs{3} , \bs{1} )_{\phantom{\text{-}}1/3, \text{-}5/3 }$ & $\blacksquare$   &            $1 \times  ( \bs{1} , \bs{1} )_{\phantom{\text{-}}1, \phantom{\text{-}}0 }$ & $\blacktriangle$  \\
\cline{1-2}
$3 \times  ( \bs{3} , \bs{1} )_{\text{-}2/3, \text{-}2/3 }$ & bulk   &             $1 \times  ( \bs{1} , \bs{1} )_{\phantom{\text{-}}1, \text{-}2 }$ & $\blacksquare$   &     $1 \times  ( \bs{1} , \bs{2} )_{\text{-}1, \text{-}1 }$ & $\bigstar$  &                 $1 \times  ( \bs{1} , \bs{1} )_{\text{-}1, \phantom{\text{-}}0 }$ & $\blacktriangle$  \\
$1 \times  ( \bs{3} , \bs{1} )_{\text{-}2/3, \phantom{\text{-}}1/3 }$ & bulk   &       $1 \times  ( \bs{1} , \bs{1} )_{\text{-}1, \phantom{\text{-}}2 }$ & $\blacksquare$   &     $1 \times  ( \bs{1} , \bs{2} )_{\text{-}1, \text{-}1 }$ & $\bullet$   &                 $1 \times  ( \bs{1} , \bs{1} )_{\phantom{\text{-}}1, \phantom{\text{-}}0 }$ & $\blacksquare$  \\
$2 \times  ( \bs{1} , \bs{2} )_{\text{-}1, \text{-}1 }$ & bulk   &                 $1 \times  ( \bs{1} , \bs{1} )_{\phantom{\text{-}}0, \phantom{\text{-}}3 }$ & $\bigstar$  & $1 \times  ( \bs{1} , \bs{2} )_{\phantom{\text{-}}0, \text{-}1 }$ & $\blacktriangle$   &           $1 \times  ( \bs{1} , \bs{1} )_{\text{-}1, \phantom{\text{-}}0 }$ & $\blacksquare$  \\
$3 \times  ( \bs{1} , \bs{2} )_{\text{-}1, \phantom{\text{-}}0 }$ & bulk   &           $1 \times  ( \bs{1} , \bs{1} )_{\phantom{\text{-}}0, \text{-}3 }$ & $\bigstar$  &      $1 \times  ( \bs{1} , \bs{2} )_{\phantom{\text{-}}0, \phantom{\text{-}}1 }$ & $\blacktriangle$   &      $1 \times  ( \bs{1} , \bs{1} )_{\phantom{\text{-}}0, \phantom{\text{-}}3 }$ & $\bigstar$ \\
$6 \times  ( \bs{1} , \bs{1} )_{\phantom{\text{-}}0, \phantom{\text{-}}2 }$ & bulk   &     $1 \times  ( \bs{1} , \bs{1} )_{\phantom{\text{-}}0, \phantom{\text{-}}3 }$ & $\bullet$   & $1 \times  ( \bs{1} , \bs{2} )_{\phantom{\text{-}}0, \phantom{\text{-}}1 }$ & $\blacksquare$   &      $1 \times  ( \bs{1} , \bs{1} )_{\phantom{\text{-}}0, \text{-}3 }$ & $\bigstar$ \\
$6 \times  ( \bs{1} , \bs{1} )_{\phantom{\text{-}}0, \text{-}2 }$ & bulk   &           $1 \times  ( \bs{1} , \bs{1} )_{\phantom{\text{-}}0, \text{-}3 }$ & $\bullet$   &      $1 \times  ( \bs{1} , \bs{2} )_{\phantom{\text{-}}0, \text{-}1 }$ & $\blacksquare$   &          $1 \times  ( \bs{1} , \bs{1} )_{\phantom{\text{-}}0, \text{-}3 }$ & $\bullet$  \\
$6 \times  ( \bs{1} , \bs{1} )_{\phantom{\text{-}}0, \text{-}1 }$ & bulk   &           $2 \times  ( \bs{1} , \bs{1} )_{\phantom{\text{-}}0, \text{-}2 }$ & $\bigstar$  &      $2 \times  ( \bs{1} , \bs{2} )_{\phantom{\text{-}}0, \phantom{\text{-}}0 }$ & $\blacktriangle$   &      $1 \times  ( \bs{1} , \bs{1} )_{\phantom{\text{-}}0, \phantom{\text{-}}3 }$ & $\bullet$  \\
$5 \times  ( \bs{1} , \bs{1} )_{\phantom{\text{-}}0, \phantom{\text{-}}1 }$ & bulk   &     $2 \times  ( \bs{1} , \bs{1} )_{\phantom{\text{-}}0, \phantom{\text{-}}2 }$ & $\bigstar$  & $2 \times  ( \bs{1} , \bs{2} )_{\phantom{\text{-}}0, \phantom{\text{-}}0 }$ & $\blacksquare$   &      $2 \times  ( \bs{1} , \bs{1} )_{\phantom{\text{-}}0, \text{-}2 }$ & $\bigstar$ \\
$16 \times ( \bs{1} , \bs{1} )_{\phantom{\text{-}}0, \text{-}1/2 }$ & bulk   &         $2 \times  ( \bs{1} , \bs{1} )_{\phantom{\text{-}}0, \text{-}2 }$ & $\bullet$   &      $1 \times  ( \bs{1} , \bs{1} )_{\phantom{\text{-}}2, \phantom{\text{-}}1 }$ & $\bigstar$  &      $2 \times  ( \bs{1} , \bs{1} )_{\phantom{\text{-}}0, \phantom{\text{-}}2 }$ & $\bigstar$ \\
$21 \times ( \bs{1} , \bs{1} )_{\phantom{\text{-}}0, \phantom{\text{-}}0 }$ & bulk   &     $2 \times  ( \bs{1} , \bs{1} )_{\phantom{\text{-}}0, \phantom{\text{-}}2 }$ & $\bullet$   & $1 \times  ( \bs{1} , \bs{1} )_{\phantom{\text{-}}2, \phantom{\text{-}}1 }$ & $\bullet$   &      $2 \times  ( \bs{1} , \bs{1} )_{\phantom{\text{-}}0, \phantom{\text{-}}2 }$ & $\bullet$  \\
\cline{1-4}
\multicolumn{2}{|c||}{$T_4$}  &                                 &   &                                         $1 \times  ( \bs{1} , \bs{1} )_{\phantom{\text{-}}1, \text{-}2 }$ & $\blacktriangle$   &            $2 \times  ( \bs{1} , \bs{1} )_{\phantom{\text{-}}0, \text{-}2 }$ & $\bullet$  \\
\cline{1-2}
$3 \times  ( \bsb{3}, \bs{1} )_{\phantom{\text{-}}2/3, \phantom{\text{-}}2/3 }$ & bulk &  & &                                             $2 \times  ( \bs{1} , \bs{1} )_{\phantom{\text{-}}1, \phantom{\text{-}}2 }$ & $\blacktriangle$   &      $3 \times  ( \bs{1} , \bs{1} )_{\phantom{\text{-}}0, \phantom{\text{-}}1 }$ & $\bigstar$ \\
$2 \times  ( \bsb{3}, \bs{1} )_{\phantom{\text{-}}2/3, \text{-}1/3 }$ & bulk&  & &                                            $2 \times  ( \bs{1} , \bs{1} )_{\text{-}1, \text{-}2 }$ & $\blacktriangle$   &                  $2 \times  ( \bs{1} , \bs{1} )_{\phantom{\text{-}}0, \text{-}1 }$ & $\bigstar$ \\
$1 \times  ( \bs{1} , \bs{2} )_{\phantom{\text{-}}1, \phantom{\text{-}}1 }$ & bulk&  &  &                                             $1 \times  ( \bs{1} , \bs{1} )_{\text{-}1, \phantom{\text{-}}2 }$ & $\blacktriangle$   &            $2 \times  ( \bs{1} , \bs{1} )_{\phantom{\text{-}}0, \text{-}1 }$ & $\bullet$  \\
$3 \times  ( \bs{1} , \bs{2} )_{\phantom{\text{-}}1, \phantom{\text{-}}0 }$ & bulk& & &                                           $2 \times  ( \bs{1} , \bs{1} )_{\phantom{\text{-}}1, \phantom{\text{-}}2 }$ & $\blacksquare$   &      $3 \times  ( \bs{1} , \bs{1} )_{\phantom{\text{-}}0, \phantom{\text{-}}1 }$ & $\bullet$  \\
$6 \times  ( \bs{1} , \bs{1} )_{\phantom{\text{-}}0, \phantom{\text{-}}2 }$ & bulk& & &                                           $1 \times  ( \bs{1} , \bs{1} )_{\phantom{\text{-}}1, \text{-}2 }$ & $\blacksquare$   &          $12 \times ( \bs{1} , \bs{1} )_{\phantom{\text{-}}0, \phantom{\text{-}}0 }$ & $\bigstar$ \\
$6 \times  ( \bs{1} , \bs{1} )_{\phantom{\text{-}}0, \text{-}2 }$ & bulk&  & &                                            $1 \times  ( \bs{1} , \bs{1} )_{\text{-}1, \phantom{\text{-}}2 }$ & $\blacksquare$   &                $12 \times ( \bs{1} , \bs{1} )_{\phantom{\text{-}}0, \phantom{\text{-}}0 }$ & $\bullet$  \\
\cline{5-8}
$4 \times  ( \bs{1} , \bs{1} )_{\phantom{\text{-}}0, \text{-}1 }$ & bulk& & &                                           &   &                                           &   \\
$6 \times  ( \bs{1} , \bs{1} )_{\phantom{\text{-}}0, \phantom{\text{-}}1 }$ & bulk& & &                                         &   &                                       &   \\
$8 \times  ( \bs{1} , \bs{1} )_{\phantom{\text{-}}0, \phantom{\text{-}}1/2 }$ & bulk& & &  &   &                                        &   \\
$12 \times ( \bs{1} , \bs{1} )_{\phantom{\text{-}}0, \phantom{\text{-}}0 }$ & bulk& & &  &   &                                      &   \\
\hline
\end{tabular}
}
\end{table}

\begin{table}[h!]
\centering \caption{The full (five dimensional) spectrum of the
models that we analyze \cite{Lebedev:2006kn}. Note that
$\mathbf{8}_{v+c+s} \equiv \mathbf{8}_{v} + \mathbf{8}_{c} +
\mathbf{8}_{s}$.  In five dimensions, both Model 1 and Model 2 have
the gauge group $\SU{6}\times\left[\SO{8}\times\SU{3}\right]'$. Note
that states are written in the language of $D=5, N=1$, and that the
spectrum of these models are identical to those examined by
Reference \cite{Buchmuller:2007qf}.}
\label{tab:6dspectrum}
\vspace{5mm} \label{full_spectrum_LNRRRVW}
\begin{footnotesize}
\begin{tabular}{c|c|c}
\hline
Multiplet Type&Representation&Number\\
\hline
\hline
tensor&singlet&1\\
\hline
vector&$(\mathbf{35},1,1)\oplus(1,\mathbf{28},1)$&35 + 28\\
&$\oplus(1,1,\mathbf{8})\oplus5\times(1,1,1)$&8 + 5\\
\hline
hyper&$(\mathbf{20},1,1)\oplus(1,\mathbf{8}_{v+c+s},1)\oplus 4\times(1,1,1)$&20+24+4 \\
&$\oplus9\times\left\{(\mbsix,1,1)\oplus(\mbsixb,1,1)\right\}$&108\\
&$\oplus9\times\left\{(1,1,\mbthree)\oplus(1,1,\mbthreeb)\right\}$&54\\
&$\oplus3\times(1,\mathbf{8}_{v+c+s},1)$&72\\
&$\oplus 36 \times (1,1,1)$&36\\
&SUGRA singlets&2\\
\hline
\end{tabular}
\end{footnotesize}
\end{table}

\begin{table}[ht!]
 \centering
 \caption{Exotic matter content in Models 1A/B and 2 from \cite{Lebedev:2007hv}.  Listed are the states' quantum numbers under the MSSM and hidden sector gauge groups, with the hypercharge denoted in the subscript.  The brane localized exotic matter in Model 1 is a subset of that in Model 2.}
 \vspace{5mm}
 \label{tab:all_exotics} \begin{footnotesize}
 \begin{tabular}{c|c|c|c|c}
 \hline
 Model&Hidden Sector&&Exotic Matter Irrep&Name\\
 \hline
 \hline
1 A/B   &$\SU{4}\times \SU2$    &brane  & $2 \times \left[(\mbthree,1;1,1)_{1/3,2/3}+(\overline{\mbthree},1;1,1)_{-1/3,-2/3}\right]$    & $v+\bar{v}$\\
    &           &exotics& $4 \times \left[(1,\mbtwo;1,1)_{0,*}+(1,\mbtwo;1,1)_{0,*}\right]$             & $m + m$\\
    &           &   & $1 \times \left[(1,\mbtwo;1,\mbtwo)_{0,0}+(1,\mbtwo;1,\mbtwo)_{0,0}\right]$           & $y + y$\\
    &           &   & $2 \times \left[(1,1;\mathbf{4},1)_{1,1}+(1,1;\overline{\mathbf{4}},1)_{-1,-1}\right]$    & $f^+ + \bar{f}^-$\\
    &           &   & $14 \times \left[(1,1;1,1)_{1,*}+(1,1;1,1)_{-1,*}\right]$                 & $s^+ + s^-$\\
\hline
    &           &bulk   &$6 \times \left[(\mbthree,1;1,1)_{-2/3,-2/3}+(\overline{\mbthree},1;1,1)_{2/3,2/3}\right] $    &$\delta + \bar{\delta}$\\
    &           &exotics&$1 \times \left[(\mbthree,1;1,1)_{-2/3,-1/3}+(\overline{\mbthree},1;1,1)_{2/3,1/3}\right] $    &$d + \bar{d}$\\
    &           &   &$1 \times\left[(1,\mbtwo;1,1)_{-1,-1}+(1,\mbtwo;1,1)_{1,1}\right]$             &$\ell + \bar{\ell}$\\
\hline
\hline
2   & $\SO8\times \SU2$     &brane  & $4 \times \left[(\mbthree,1;1,1)_{1/3,*}+(\overline{\mbthree},1;1,1)_{-1/3,*}\right] $    & $v+\bar{v}$\\
    &           &exotics& $2 \times \left[(1,\mbtwo;1,1)_{0,*}+(1,\mbtwo;1,1)_{0,*}\right]$             & $m + m$\\
    &           &   & $1 \times \left[(1,\mbtwo;1,\mbtwo)_{0,0}+(1,\mbtwo;1,\mbtwo)_{0,0}\right]$           & $y + y$\\
    &           &   & $2 \times \left[(1,1;1,\mbtwo)_{1,1}+(1,1;1,\mbtwo)_{-1,-1}\right]$               & $x^+ + x^-$\\
    &           &   & $20 \times \left[(1,1;1,1)_{1,*}+(1,1;1,1)_{-1,*}\right]$                 & $s^+ + s^-$\\
\hline
    &           &bulk   & $3 \times \left[(\mbthree,1;1,1)_{-2/3-2/3}+(\overline{\mbthree},1;1,1)_{2/3,2/3}\right]$ &$\delta + \bar{\delta}$\\
    &           &exotics&$1 \times \left[(\mbthree,1;1,1)_{-2/3,2/3}+(\overline{\mbthree},1;1,1)_{2/3,-2/3}\right]$ &$d + \bar{d}$\\
    &           &   &$1 \times\left[(1,\mbtwo;1,1)_{-1,-1}+(1,\mbtwo;1,1)_{1,1}\right]$             &$\ell + \bar{\ell}$\\
    &           &   &$3 \times\left[(1,\mbtwo;1,1)_{-1,0}+(1,\mbtwo;1,1)_{1,0}\right]$              &$\phi + \bar{\phi}$\\
 \hline
 \end{tabular}
 \end{footnotesize}
\end{table}

\begin{table}[ht!]
 \centering
 \caption{Values of the $\beta$-function coefficients for the \textit{brane-localized} exotic matter.  These states do not have zero modes, and come from the $T^3$ and $T^1/T^5$ sectors of the theory.}
 \vspace{5mm}
 \label{tab:vector_like_brane} \begin{footnotesize}
 \begin{tabular}{c|c|c|c|c}
 \hline
 irrep&Mult (Model 2)&$b_3$&$b_2$&$b_Y$\\
 \hline
 \hline
 $ (\mbthree,1)_{1/3}+(\overline{\mbthree},1)_{-1/3} $&4&1&0&1/10\\
 $ (1,\mbtwo)_0+(1,\mbtwo)_0 $&4&0&1&0\\
 $ (1,1)_{1}+(1,1)_{-1} $&24&0&0&3/10\\
 \hline
 \end{tabular}
 \end{footnotesize}
\end{table}

\begin{table}[ht!]
 \centering
 \caption{Values of the $\beta$-function coefficients for matter living in the bulk, along with their embeddings into \SU{6}.  (The group branching rules for $\SU{6} \rightarrow \SU{5}\times \U{1}$can be found in Reference \cite{Slansky:1981yr}.)  It is important to distinguish whether these are vector (V) or chiral (C) multiplets.}
 \vspace{5mm}
 \label{tab:all_b_values} \begin{footnotesize}
 \begin{tabular}{c|c|c|c|c|c||c|c|c|c|c}
 \hline
 \SU{6} rep&&irrep&$b_3^{++}$&$b_2^{++}$&$b_Y^{++}$&&irrep&$b_3^{--}$&$b_2^{--}$&$b_Y^{--}$\\
 \hline
 \hline
 &V&$(\mathbf{8},1)_0$&-9&0&0&C&$(\mathbf{8},1)_0$&3&0&0\\
 $\mathbf{35}$&V&$(1,\mbthree)_0$&0&-6&0&C&$(1,\mbthree)_0$&0&2&0\\
 &C&$(1,\mbtwo)_{1}$&0&1/2&3/10&V&$(1,\mbtwo)_{-1}$&0&-3/2&-9/10\\
 &C&$(1,\mbtwo)_{-1}$&0&1/2&3/10&V&$(1,\mbtwo)_{1}$&0&-3/2&-9/10\\
 \hline
 &C&$(\mbthree,\mbtwo)_{1/3}$&1&3/2&1/10&C& $(\overline{\mbthree},\mbtwo)_{-1/3}$&1&3/2&1/10\\
 $\mathbf{20}$&C&$(\overline{\mbthree},1)_{-4/3}$&1/2&0&4/5&C&$(\mbthree,1)_{4/3}$&1/2&0&4/5\\
 &C&$(1,1)_{2}$&0&0&3/5&C&$(1,1)_{-2}$&0&0&3/5\\
 \hline
 $\mbsix + \mbsixb$&C&$(1,\mbtwo)_{-1}$&0&1/2&3/10&C&$(1,\mbtwo)_{1}$&0&1/2&3/10\\
 &C&$(\overline{\mbthree},1)_{2/3}$&1/2&0&1/5&C&$(\mbthree,1)_{-2/3}$&1/2&0&1/5\\
 \hline
\end{tabular}
\end{footnotesize}
\end{table}

\begin{table}[ht!]
\renewcommand{\arraystretch}{1.2}
 \centering
 \caption{Comparison of proton lifetime to $M_{\st}$, $\mc$, and $\mex$, in the case where
 no exotic matter lives in the bulk.  In general, an intermediate scale is needed to fit the
 low energy data and the proton decay constraints.  We have used $\beta_{\textsc{lattice}} \simeq 0.011$
 \cite{Aoki:2006ib}.  We note the solutions which will also work for Model 1A in \textbf{bold}.
 Note that $\vec{n}$ refers to \textit{brane localized} exotics only, and is defined in
 Equation (\ref{define_n_values}).}
 \vspace{5mm}
 \label{tab:proton_decay}

 \begin{tabular}{c|c|c|c|c}
  \hline
  $\vec{n}$&$M_{\st}$ in GeV & $\mc$ in GeV & $\mex$ in GeV &$\tau(p\rightarrow e^+ \pi^0)$ in yr\\
  \hline
  \hline
$\mathbf{(2,1,0)}$&$\mathbf{9.18\times 10^{ 17}}$&$\mathbf{2.22\times 10^{ 17}}$&$\mathbf{2.60\times 10^{ 9}}$&$\mathbf{1.77\times 10^{ 38}}$\\
$(4,2,0)$&$9.18\times 10^{ 17}$&$2.22\times 10^{ 17}$&$4.88\times 10^{ 13}$&$1.77\times 10^{ 38}$\\
$(3,2,3)$&$9.88\times 10^{ 17}$&$2.22\times 10^{ 17}$&$2.08\times 10^{ 9}$&$1.32\times 10^{ 38}$\\
$(4,3,6)$&$1.08\times 10^{ 18}$&$2.22\times 10^{ 17}$&$1.59\times 10^{ 9}$&$9.23\times 10^{ 37}$\\
$(4,2,1)$&$8.26\times 10^{ 17}$&$6.65\times 10^{ 16}$&$5.43\times 10^{ 13}$&$2.19\times 10^{ 36}$\\
$(4,2,2)$&$6.87\times 10^{ 17}$&$2.19\times 10^{ 16}$&$6.52\times 10^{ 13}$&$5.34\times 10^{ 34}$\\
$\mathbf{(2,1,1)}$&$\mathbf{6.87\times 10^{ 17}}$&$\mathbf{2.19\times 10^{ 16}}$&$\mathbf{6.18\times 10^{ 9}}$&$\mathbf{5.34\times 10^{ 34}}$\\
$(3,2,4)$&$7.07\times 10^{ 17}$&$2.16\times 10^{ 16}$&$5.68\times 10^{ 9}$&$4.52\times 10^{ 34}$\\
$(4,3,7)$&$7.28\times 10^{ 17}$&$2.13\times 10^{ 16}$&$5.21\times 10^{ 9}$&$3.79\times 10^{ 34}$\\
$(3,1,0)$&$5.43\times 10^{ 17}$&$8.20\times 10^{ 15}$&$8.25\times 10^{ 13}$&$2.70\times 10^{ 33}$\\
$(4,2,3)$&$5.47\times 10^{ 17}$&$8.15\times 10^{ 15}$&$8.19\times 10^{ 13}$&$2.57\times 10^{ 33}$\\
\hline
 \end{tabular}

\end{table}

\begin{table}
\renewcommand{\arraystretch}{1.4}
 \centering
 \caption{Comparison of proton lifetime to $M_{\st}$, $\mc$, and $\mex$.  In general,
 an intermediate scale is needed to fit the low energy data and the proton decay constraints.
 We have used $\beta_{\textsc{lattice}}\simeq 0.011$ \cite{Aoki:2006ib}.  Note that $\vec{n}$ refers
 to \textit{brane localized} exotics only, and is defined in Equation (\ref{define_n_values}).
 For details on the solution marked with an arrow ($\Rightarrow$), see Section \ref{sec:top_down}.
 We note the solutions which will also work for Model 1A in \textbf{bold}.}
 \vspace{5mm}
 \label{tab:proton_decay_two}
 \scriptsize
 \begin{tabular}{c|c|c|c|c|c}
\hline
  Bulk Exotics&$\vec{n}$&$M_{\st}$ in GeV & $\mc$ in GeV & $\mex$ in GeV & $\tau(p\rightarrow e^+ \pi^0)$ in yr\\
\hline
\hline
$\left[(\mbthree,1)_{2/3,*} + (\mbthreeb,1)_{-2/3,*})\right]^{++}$ +  &$(4,3,1)$&$9.96\times 10^{17}$&$7.74\times 10^{17}$&$4.50\times 10^{13}$&$1.90\times 10^{40}$\\
$\left[(1,\mbtwo)_{1,*} + (1,\mbtwo)_{-1,*})\right]^{--}$  &$(4,3,2)$&$9.73\times 10^{17}$&$2.22\times 10^{17}$&$4.61\times 10^{13}$&$1.40\times 10^{38}$\\
&$\Rightarrow\mathbf{(2,2,2)}$&$\mathbf{1.01\times 10^{18}}$&$\mathbf{2.22\times 10^{17}}$&$\mathbf{1.92\times 10^{ 9}}$&$\mathbf{1.19\times 10^{38}}$\\
&$(3,3,5)$&$1.12\times 10^{18}$&$2.22\times 10^{17}$&$1.43\times 10^{ 9}$&$7.97\times 10^{37}$\\
&$(4,4,8)$&$1.28\times 10^{18}$&$2.22\times 10^{17}$&$9.64\times 10^{ 8}$&$4.73\times 10^{37}$\\
&$(3,2,0)$&$8.79\times 10^{17}$&$6.55\times 10^{16}$&$5.10\times 10^{13}$&$1.61\times 10^{36}$\\
&$(4,3,3)$&$9.06\times 10^{17}$&$6.50\times 10^{16}$&$4.95\times 10^{13}$&$1.38\times 10^{36}$\\
&$(3,2,1)$&$7.67\times 10^{17}$&$2.07\times 10^{16}$&$5.84\times 10^{13}$&$2.77\times 10^{34}$\\
&$\mathbf{(1,1,0)}$&$\mathbf{7.67\times 10^{17}}$&$\mathbf{2.07\times 10^{16}}$&$\mathbf{4.45\times 10^{ 9}}$&$\mathbf{2.77\times 10^{34}}$\\
&$(4,3,4)$&$7.82\times 10^{17}$&$2.05\times 10^{16}$&$5.73\times 10^{13}$&$2.47\times 10^{34}$\\
&$\mathbf{(2,2,3)}$&$\mathbf{7.97\times 10^{17}}$&$\mathbf{2.03\times 10^{16}}$&$\mathbf{3.96\times 10^{ 9}}$&$\mathbf{2.20\times 10^{34}}$\\
&$(3,3,6)$&$8.31\times 10^{17}$&$1.99\times 10^{16}$&$3.50\times 10^{ 9}$&$1.71\times 10^{34}$\\
&$(4,4,9)$&$8.69\times 10^{17}$&$1.95\times 10^{16}$&$3.06\times 10^{ 9}$&$1.31\times 10^{34}$\\
&$(4,2,0)$&$6.69\times 10^{17}$&$1.03\times 10^{16}$&$1.44\times 10^{15}$&$2.92\times 10^{33}$\\
\hline
$\left[(\mbthree,1)_{2/3,*} + (\mbthreeb,1)_{-2/3,*})\right]^{--}$ +  &$(3,1,1)$&$1.01\times 10^{18}$&$2.22\times 10^{17}$&$1.92\times 10^{ 9}$&$1.19\times 10^{38}$\\
$\left[(1,\mbtwo)_{1,*} + (1,\mbtwo)_{-1,*})\right]^{++}$  &$(4,2,4)$&$1.12\times 10^{18}$&$2.22\times 10^{17}$&$1.43\times 10^{ 9}$&$7.97\times 10^{37}$\\
&$(4,1,0)$&$7.67\times 10^{17}$&$2.07\times 10^{16}$&$5.84\times 10^{13}$&$2.77\times 10^{34}$\\
&$(3,1,2)$&$7.97\times 10^{17}$&$2.03\times 10^{16}$&$3.96\times 10^{ 9}$&$2.20\times 10^{34}$\\
&$(4,2,5)$&$8.31\times 10^{17}$&$1.99\times 10^{16}$&$3.50\times 10^{ 9}$&$1.71\times 10^{34}$\\
\hline
$\left[(\mbthree,1)_{2/3,*} + (\mbthreeb,1)_{-2/3,*})\right]^{++}$ +  &$\mathbf{(2,1,0)}$&$\mathbf{1.01\times 10^{18}}$&$\mathbf{2.22\times 10^{17}}$&$\mathbf{1.92\times 10^{ 9}}$&$\mathbf{1.19\times 10^{38}}$\\
$\left[(1,\mbtwo)_{1,*} + (1,\mbtwo)_{-1,*})\right]^{++}$ &$(3,2,3)$&$1.12\times 10^{18}$&$2.22\times 10^{17}$&$1.43\times 10^{ 9}$&$7.97\times 10^{37}$\\
&$(4,2,0)$&$9.73\times 10^{17}$&$2.22\times 10^{17}$&$4.61\times 10^{13}$&$1.40\times 10^{38}$\\
&$(4,3,6)$&$1.28\times 10^{18}$&$2.22\times 10^{17}$&$9.64\times 10^{ 8}$&$4.73\times 10^{37}$\\
&$(4,2,1)$&$9.06\times 10^{17}$&$6.50\times 10^{16}$&$4.95\times 10^{13}$&$1.38\times 10^{36}$\\
&$(4,2,2)$&$7.82\times 10^{17}$&$2.05\times 10^{16}$&$5.73\times 10^{13}$&$2.47\times 10^{34}$\\
&$\mathbf{(2,1,1)}$&$\mathbf{7.97\times 10^{17}}$&$\mathbf{2.03\times 10^{16}}$&$\mathbf{3.96\times 10^{ 9}}$&$\mathbf{2.20\times 10^{34}}$\\
&$(3,2,4)$&$8.31\times 10^{17}$&$1.99\times 10^{16}$&$3.50\times 10^{ 9}$&$1.71\times 10^{34}$\\
&$(4,3,7)$&$8.69\times 10^{17}$&$1.95\times 10^{16}$&$3.06\times 10^{ 9}$&$1.31\times 10^{34}$\\
\hline
$\left[(\mbthree,1)_{2/3,*} + (\mbthreeb,1)_{-2/3,*})\right]^{--}$ +  &$\mathbf{(2,1,0)}$&$\mathbf{9.36\times 10^{17}}$&$\mathbf{2.22\times 10^{17}}$&$\mathbf{2.45\times 10^{ 9}}$&$\mathbf{1.64\times 10^{38}}$\\
$\left[(1,\mbtwo)_{1,*} + (1,\mbtwo)_{-1,*})\right]^{--}$ &$(4,2,0)$&$9.36\times 10^{17}$&$2.22\times 10^{17}$&$4.79\times 10^{13}$&$1.64\times 10^{38}$\\
&$(3,2,3)$&$1.01\times 10^{18}$&$2.22\times 10^{17}$&$1.92\times 10^{ 9}$&$1.19\times 10^{38}$\\
&$(4,3,6)$&$1.12\times 10^{18}$&$2.22\times 10^{17}$&$1.43\times 10^{ 9}$&$7.97\times 10^{37}$\\
&$(4,2,1)$&$8.79\times 10^{17}$&$6.55\times 10^{16}$&$5.10\times 10^{13}$&$1.61\times 10^{36}$\\
&$\mathbf{(2,1,1)}$&$\mathbf{7.67\times 10^{17}}$&$\mathbf{2.07\times 10^{16}}$&$\mathbf{4.45\times 10^{ 9}}$&$\mathbf{2.77\times 10^{34}}$\\
&$(4,2,2)$&$7.67\times 10^{17}$&$2.07\times 10^{16}$&$5.84\times 10^{13}$&$2.77\times 10^{34}$\\
&$(3,2,4)$&$7.97\times 10^{17}$&$2.03\times 10^{16}$&$3.96\times 10^{ 9}$&$2.20\times 10^{34}$\\
&$(4,3,7)$&$8.31\times 10^{17}$&$1.99\times 10^{16}$&$3.50\times 10^{ 9}$&$1.71\times 10^{34}$\\
\hline
 \end{tabular}
\normalsize
\end{table}

\begin{table}[ht!]
\renewcommand{\arraystretch}{1.2}
 \centering
 \caption{Subset of models listed in Tables \ref{tab:proton_decay} and \ref{tab:proton_decay_two} which exhibit moderate hierarchies between all of the scales in the problem, as pictured in Figure \ref{fig:scatter}, in the red box.  Note that none of these results can be accommodated in Model 1A.}
 \vspace{5mm}
 \label{tab:interesting_models}
 \scriptsize
 \begin{tabular}{c|c|c|c|c|c}
  \hline
  Bulk Exotics & $\vec{n}$&$M_{\st}$ in GeV & $\mc$ in GeV & $\mex$ in GeV &$\tau(p\rightarrow e^+ \pi^0)$ in yr\\
  \hline
  \hline
  None&$(4,2,3)$&$5.47\times 10^{ 17}$&$8.15\times 10^{ 15}$&$8.19\times 10^{ 13}$&$2.57\times 10^{ 33}$\\
  &$(3,1,0)$&$5.43\times 10^{ 17}$&$8.20\times 10^{ 15}$&$8.25\times 10^{ 13}$&$2.70\times 10^{ 33}$\\
  &$(4,2,2)$&$6.87\times 10^{ 17}$&$2.19\times 10^{ 16}$&$6.52\times 10^{ 13}$&$5.34\times 10^{ 34}$\\
  \hline
  $\left[(\mbthree,1)_{2/3,*} + (\mbthreeb,1)_{-2/3,*})\right]^{++}$ +  &$(4,2,0)$&$6.69\times 10^{17}$&$1.03\times 10^{16}$&$1.44\times 10^{15}$&$2.92\times 10^{33}$\\
  $\left[(1,\mbtwo)_{1,*} + (1,\mbtwo)_{-1,*})\right]^{--}$  &$(4,3,4)$&$7.82\times 10^{17}$&$2.05\times 10^{16}$&$5.73\times 10^{13}$&$2.47\times 10^{34}$\\
  &$(3,2,1)$&$7.67\times 10^{17}$&$2.07\times 10^{16}$&$5.84\times 10^{13}$&$2.77\times 10^{34}$\\
  \hline
  $\left[(\mbthree,1)_{2/3,*} + (\mbthreeb,1)_{-2/3,*})\right]^{--}$ + &$(4,1,0)$&$7.67\times 10^{17}$&$2.07\times 10^{16}$&$5.84\times 10^{13}$&$2.77\times 10^{34}$\\
  $\left[(1,\mbtwo)_{1,*} + (1,\mbtwo)_{-1,*})\right]^{++}$&&&&&\\
  \hline
  $\left[(\mbthree,1)_{2/3,*} + (\mbthreeb,1)_{-2/3,*})\right]^{++}$ +  &$(4,2,2)$&$7.82\times 10^{17}$&$2.05\times 10^{16}$&$5.73\times 10^{13}$&$2.47\times 10^{34}$\\
  $\left[(1,\mbtwo)_{1} + (1,\mbtwo)_{-1})\right]^{++}$&&&&&\\

  \hline
  $\left[(\mbthree,1)_{2/3,*} + (\mbthreeb,1)_{-2/3,*})\right]^{--}$ +  &$(4,2,2)$&$7.67\times 10^{17}$&$2.07\times 10^{16}$&$5.84\times 10^{13}$&$2.77\times 10^{34}$\\
  $\left[(1,\mbtwo)_{1} + (1,\mbtwo)_{-1})\right]^{--}$ &&&&&\\

  \hline
  \hline
 \end{tabular}
\normalsize
\end{table}

\clearpage
\bibliographystyle{utphys}
\bibliography{mybibliography}

\end{document}